\documentclass[envcountsame,runningheads,notitlepage]{llncs}

\usepackage{algorithm}
\usepackage{algpseudocode}
\usepackage{algorithmicx}
\usepackage{listings}
\usepackage{color}
\usepackage{amsmath}  
\usepackage{graphicx} 
\usepackage{amssymb} 
\usepackage{algpseudocode}
\usepackage{algorithm}
\usepackage{comment}
\usepackage{hyperref}
\usepackage{orcidlink}
\usepackage{pgfplots}
\usepackage[english]{babel}
\usepackage{float}
\floatstyle{plaintop}
\restylefloat{table}
\usepackage{pythonhighlight}
\usepackage{lscape}

\usepackage{graphicx}
\usepackage{epstopdf}
\usepackage{algorithm}
\usepackage{algpseudocode}
\AppendGraphicsExtensions{.gif}

\begin{document}

\title{Evaluation of Privacy-aware Support Vector Machine (SVM) Learning using Homomorphic Encryption}
\titlerunning{Evaluation of Privacy-aware SVM}

\author{
 William J Buchanan\inst{1} \orcidlink{0000-0003-0809-3523} \and Hisham Ali \inst{1} \orcidlink{0000-0002-0333-4757}}

\institute{Blockpass ID Lab, Edinburgh Napier University\\
  \href{mailto:b.buchanan@napier.ac.uk}{b.buchanan@napier.ac.uk} 
}  %

\maketitle

\begin{abstract}
The requirement for privacy-aware machine learning increases as we continue to use PII (Personally Identifiable Information) within machine training. To overcome these privacy issues, we can apply Fully Homomorphic Encryption (FHE) to encrypt data before it is fed into a machine learning model. This involves creating a homomorphic encryption key pair, and where the associated public key will be used to encrypt the input data, and the private key will decrypt the output. But, there is often a performance hit when we use homomorphic encryption, and so this paper evaluates the performance overhead of using the SVM machine learning technique with the OpenFHE homomorphic encryption library. This uses Python and the scikit-learn library for its implementation.  The experiments include a range of variables such as multiplication depth, scale size, first modulus size, security level, batch size, and ring dimension, along with two different SVM models, SVM-Poly and SVM-Linear. Overall, the results show that the two main parameters which affect performance are the ring dimension and the modulus size, and that SVM-Poly and SVM-Linear show similar performance levels.
\end{abstract}


\section{Introduction}
The rise in machine learning (ML) has caused an increasing demand for data to be used in creating data for learning. Unfortunately, this data can also include PII, and where it is often needed to be protected before it is shared. While data can be protected \emph{over-the-air} and \emph{at-rest}, we often do not protect data \emph{in-process}. For this, we can use homomorphic encryption to process encrypted data. This can either be Partial Homomorphic Encryption (PHE) or Fully Homomorphic Encryption (FHE). With FHE, we can implement all of the arithmetic operations, while PHE only implements a reduced number of operations. With this, FHE typically uses lattice cryptography, and which often has an increased processing requirement for its implementation. This paper thus makes a core contribution in applying FHE to the SVM (Support Vector Machine) models, and then evaluates the performance of this using a range of parameters using within the OpenFHE library \cite{openfhe_github}. 

\section{Background}
Homomorphic encryption supports mathematical operations on encrypted data. In 1978, Rivest, Adleman, and Dertouzos \cite{rivest1978data} were the first to define the possibilities of implementing a homomorphic operation and used the RSA method. This supported multiply and divide operations \cite{asecuritysite_17070}, but does not support addition and subtraction. Overall, PHE supports a few arithmetic operations, while FHE supports add, subtract, multiply, and divide. 

Since Gentry defined the first FHE method \cite{homenc} in 2009, there have been four main generations of homomorphic encryption:

\begin{itemize}
    \item 1st generation: Gentry’s method uses integers and lattices \cite{van2010fully} including the DGHV method.
    \item 2nd generation. Brakerski, Gentry and Vaikuntanathan’s (BGV) and Brakerski/ Fan-Vercauteren (BFV) use a Ring Learning With Errors approach \cite{brakerski2014efficient}.  The methods are similar to each other, and there is only a small difference between them.
    \item 3rd generation: These include DM (also known as FHEW) and CGGI (also known as TFHE) and support the integration of  Boolean circuits for small integers. 
    \item 4th generation: CKKS (Cheon, Kim, Kim, Song) and which uses floating-point numbers \cite{cheon2017homomorphic}.
\end{itemize}

Generally, CKKS works best for real number computations and can be applied to machine learning applications as it can implement logistic regression methods and other statistical computations. DM (also known as FHEW) and CGGI (also known as TFHE) are useful in the application of Boolean circuits for small integers. BGV and BFV are generally used in applications with small integer values.

\subsection{Public key or symmetric key}
\begin{figure}
  \includegraphics[width=\linewidth]{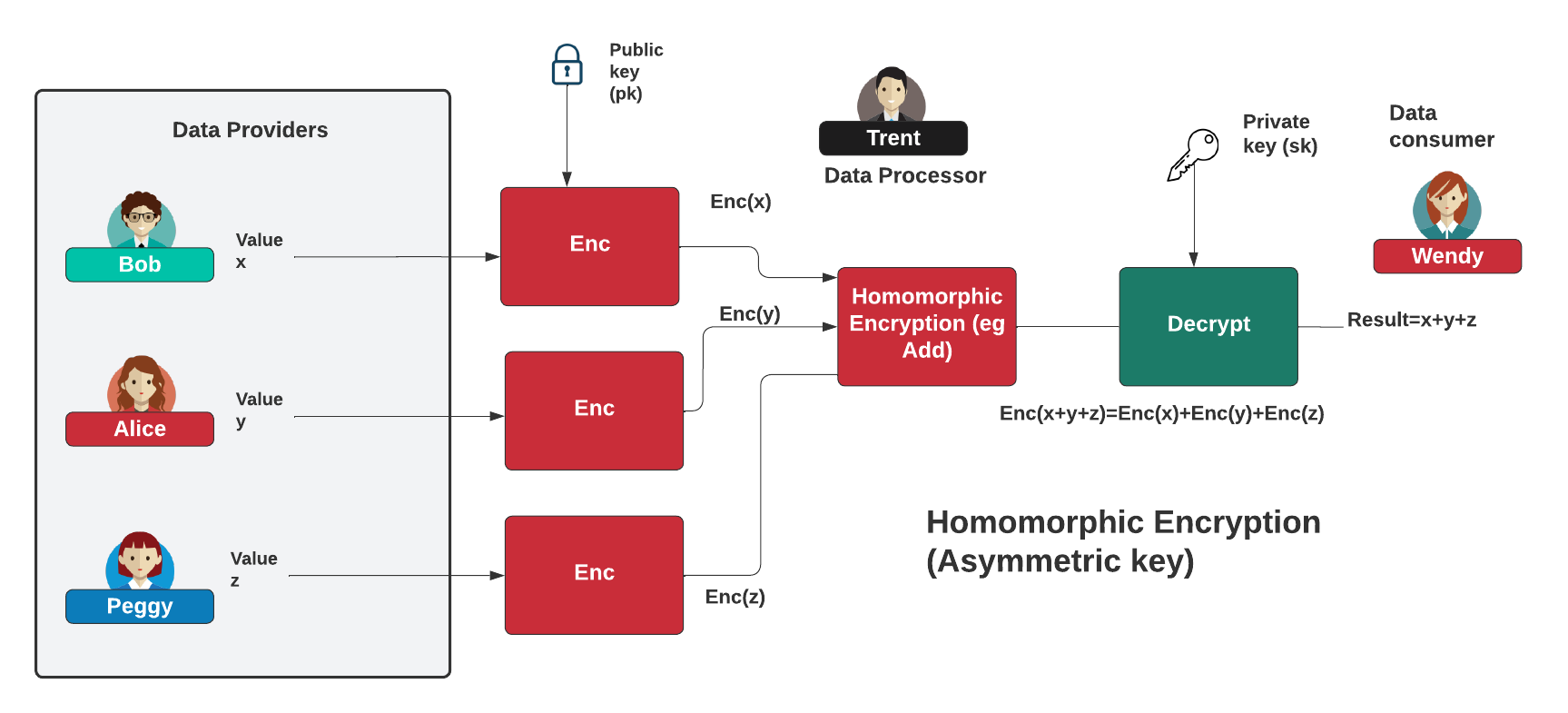}
  \caption{Asymmetric encryption (public key)}
  \label{fig:asym}
\end{figure}

\begin{figure}
  \includegraphics[width=\linewidth]{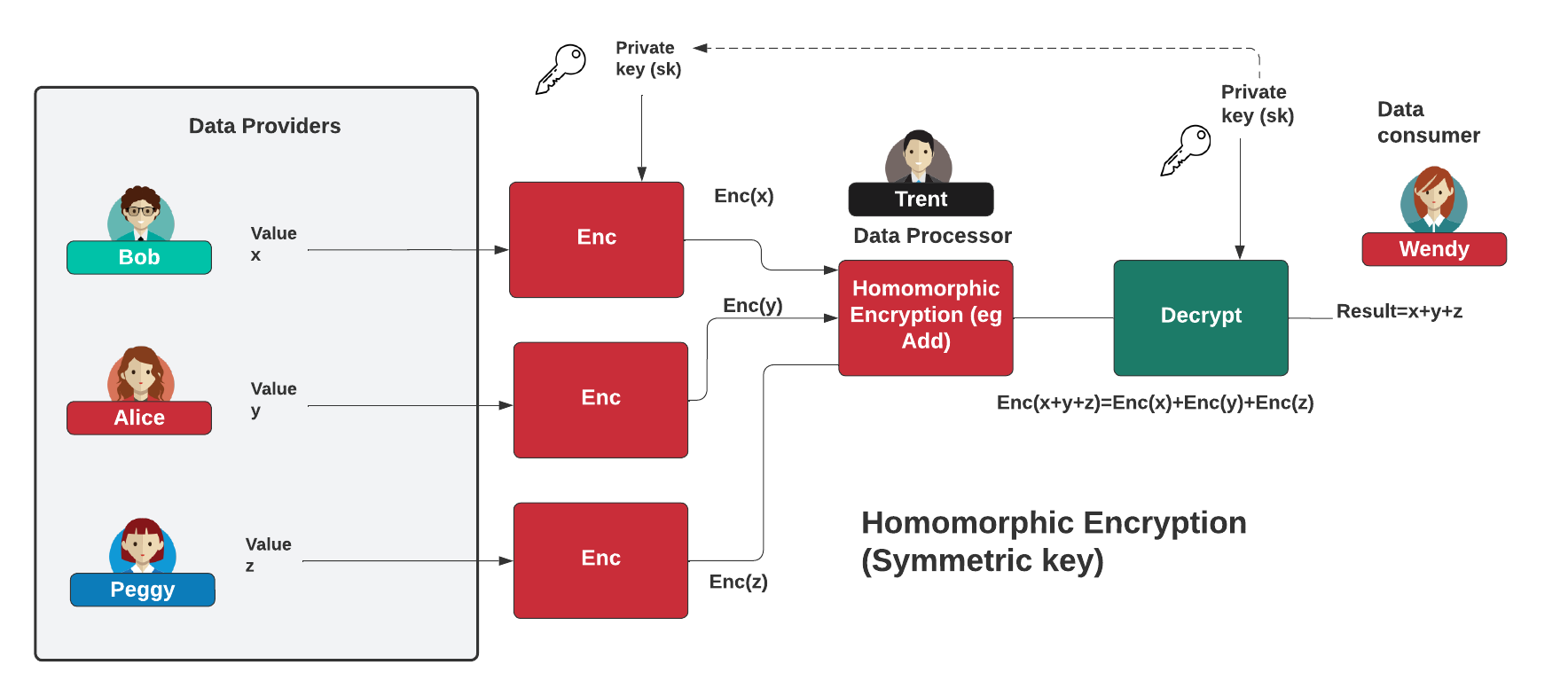}
  \caption{Symmetric encryption}
  \label{fig:sym}
\end{figure}

Homomorphic encryption can be implemented either with a symmetric key or an asymmetric (public) key. With symmetric key encryption, we use the same key to encrypt as we do to decrypt, whereas, with an asymmetric method, we use a public key to encrypt and a private key to decrypt.  In Figure \ref{fig:asym} we use asymmetric encryption with a public key ($pk$) and a private key ($sk$). With this Bob, Alice and Peggy will encrypt their data using the public key to produce ciphertext, and then we can operate on the ciphertext using arithmetic operations. The result can then be revealed by decrypting with the associated private key. In Figure \ref{fig:sym} we use symmetric key encryption, and where the data is encrypted with a secret key, and which is then used to decrypt the data. In this case, the data processor (Trent) should not have access to the secret key, as they could decrypt the data from the providers.

\subsection{Homomorphic libraries}
There are several homomorphic encryption libraries that support FHE, including ones that support CUDA and GPU acceleration, but many have not been kept up-to-date with modern methods or have only integrated one method. Overall, the native language libraries tend to be the most useful, as they allow the compilation to machine code. The main languages used for this are C++, Golang, and Rust, although some Python libraries exist through wrappers of C++ code. This includes HEAAN-Python, and its associated HEAAN library.  

One of the first libraries which supported a range of methods is Microsoft SEAL \cite{asecuritysite_85691}, SEAL-C\# and SEAL-Python. While it supports a wide range of methods, including BGV/BFV and CKKS, it has lacked any real serious development for the past few years. It does have support for Android and has a Node.js port \cite{asecuritysite_40933}. Wood et al. \cite{wood2020homomorphic} define a full range of libraries. One of the most extensive libraries is PALISADE, and which has now developed into OpenFHE. Within OpenFHE. The main implementations is this library are:

\begin{itemize}
    \item Brakerski/Fan-Vercauteren (\textbf{BFV}) scheme for integer arithmetic
    \item Brakerski-Gentry-Vaikuntanathan (\textbf{BGV}) scheme for integer arithmetic
    \item Cheon-Kim-Kim-Song (\textbf{CKKS}) scheme for real-number arithmetic (includes approximate bootstrapping)
    \item Ducas-Micciancio (\textbf{DM}) and Chillotti-Gama-Georgieva-Izabachene (\textbf{CGGI}) schemes for Boolean circuit evaluation.
\end{itemize}

\subsection{Bootstrapping}
\begin{figure}
  \includegraphics[width=\linewidth]{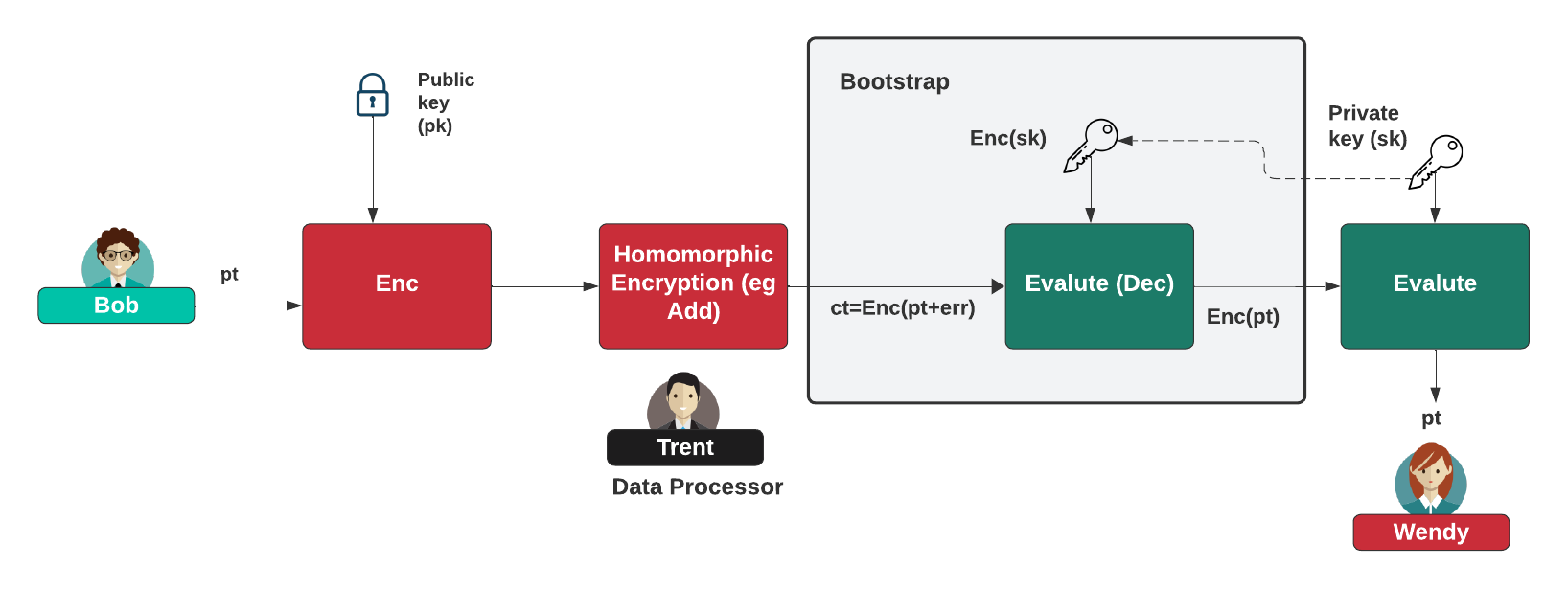}
  \caption{Bootstrap}
  \label{fig:bootstrap}
\end{figure}

A key topic within fully homomorphic encryption is the usage of bootstrapping. Within a learning with-errors approach, we add noise to our computations. For a normal decryption process, we use the public key to encrypt data and then the associated private key to decrypt it. Within the bootstrap version of homomorphic encryption, we use an encrypted version of the private key that operates on the ciphertext. In this way, we remove the noise which can build up in the computation. Figure \ref{fig:bootstrap} outlines that we perform an evaluation on the decryption using an encrypted version of the private key. This will remove noise in the ciphertext, after which we can then use the actual private key to perform the decryption.

The main bootstrapping methods are CKKS \cite{cheon2017homomorphic}, DM \cite{ducas2015fhew}/CGGI, and BGV/BFV. Overall, CKKS is generally the fastest bootstrapping method, while DM/CGGI is efficient with the evaluation of arbitrary functions. These functions approximate math functions as polynomials (such as with  Chebyshev approximation). BGV/BFV provides reasonable performance and is generally faster than DM/CGGI but slower than CKKS.

\subsection{Arbitrary smooth functions}
With approximation theory, it is possible to determine an approximate polynomial $p(x)$ that is an approximation to a function $f(x)$. A polynomial takes the form of $p(x)=a_n.x^n+a_{n-1}.x^{n-1}+...+a_1.x+a_0$, and where $a_0$... $a_n$ are the coefficients of the powers, and $n$ is the maximum power of the polynomial. 

For this, we can define arbitrary smooth functions for CKKS using Chebyshev approximation \cite{al2023demystifying}. These were initially created by Pafnuty Lvovich Chebyshev. This method involves the approximation of a smooth function using polynomials. Examples of these functions include $log_{10}$, $log_2$, $log_e$, and $e^x$  \cite{asecuritysite_65179}.

\subsection{Plaintext slots}
\begin{figure}
\centering
  \includegraphics[width=0.6\linewidth]{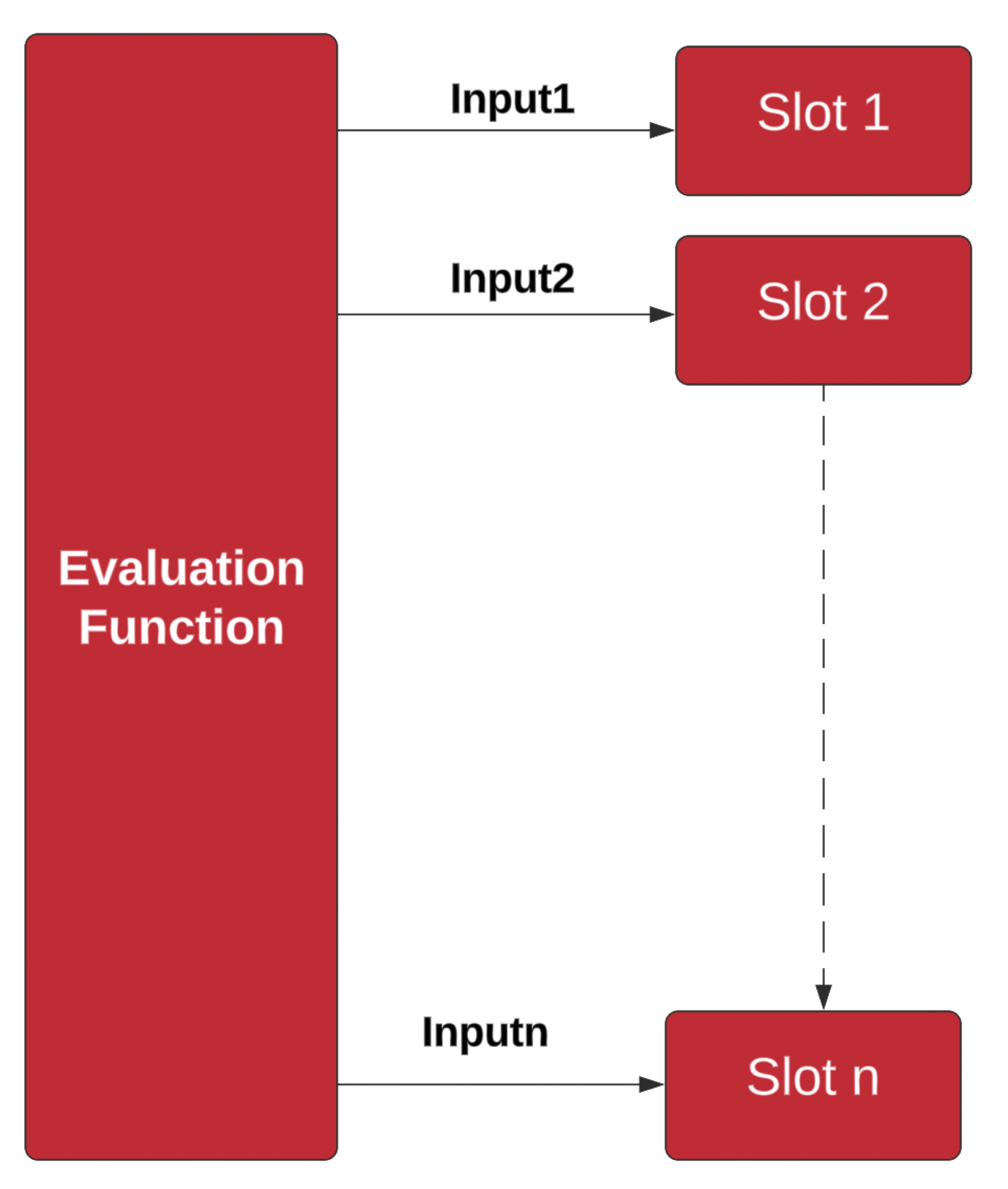}
  \caption{Slots for plaintext}
  \label{fig:slots}
\end{figure}

With many homomorphic methods, we can encrypt multiple plaintext values into ciphertext in a single operation. This is defined as the number of plaintext slots, and is illustrated in Figure \ref{fig:slots}.

\subsection{BGV and BFV}
With BGV and BFV, we use a Ring Learning With Errors (LWE) method \cite{brakerski2014efficient}.  With BGV, we define a moduli ($q$), which constrains the range of the polynomial coefficients. Overall, the methods use a moduli, which can be defined within different levels. We then initially define a finite group of $\mathbb{Z}_q$, and then make this a ring by dividing our operations with $(x^n+1)$ and where $n-1$ is the largest power of the coefficients. The message can then be represented in binary as:

\begin{equation}
m=a_{n-1}a_{n-2}...a_0
\end{equation}

This can be converted into a polynomial with:

\begin{equation}
\mathbf{m}=a_{n-1} x^{n-1} + a_{n-2} x^{n-2}+...+a_1 x + a_0 \pmod q
\end{equation}

The coefficients of this polynomial will then be a vector. Note that for efficiency, we can also encode the message with ternary (such as with -1, 0 and 1). We then define the plaintext modulus with:

\begin{equation}
t = p^r
\end{equation}

and where $p$ is a prime number and $r$ is a positive number. We can then define a ciphertext modulus of $q$, and which should be much larger than $t$. To encrypt with the private key of $\mathbf{s}$, we implement:

\begin{equation}
(c_0, c_1) =\left( \frac{q}{t}.\mathbf{m}  + \mathbf{a}.\mathbf{s} + e,\mathbf{-a} \right) \mod q
\end{equation}

To decrypt:

\begin{equation}
m = \bigl \lfloor \frac{t}{q}(c_0+c_1).\mathbf{s} \bigr \rceil
\end{equation}

This works because:

\begin{align}
m_{recover} &= \bigl \lfloor  \frac{t}{q}\left(\frac{q}{t}.\mathbf{m}  + \mathbf{a}.\mathbf{s} + e -\mathbf{a}.\mathbf{s} \right) \bigr \rceil\\
&= \bigl \lfloor \left( \mathbf{m}  + \frac{t}{q}.e  \right) 
 \bigr \rceil\\
 & \approx m \\
\end{align}

For two message of $m_1$ and $m_2$, we will get:

\begin{align}
Enc(m_1+m_2) &= Enc(m_1) + Enc(m_2)\\
Enc(m_1.m_2) &= Enc(m_1) . Enc(m_2)
\end{align}

\subsubsection{Noise and computation}
But each time we add or multiply, the error also increases. Thus bootstrapping is required to reduce the noise. Overall, addition and plaintext/ciphertext multiplication is not a time-consuming task, but ciphertext/ciphertext multiplication is more computationally intensive. The most computational task is typically the bootstrapping process, and the ciphertext/ciphertext multiplication process adds the most noise to the process.

\subsubsection{Parameters}
We thus have a parameter of the ciphertext modulus (q) and the plaintext modulus (t). Both of these are typically to the power of 2. An example of $q$ is $2^{240}$ and for $t$ is 65,537. As the value of $2^q$ is likely to be a large number, we typically define it as a $log\_q$ value.  Thus, a ciphertext modulus of $2^{240}$ will be 240 as defined as a $log_q$ value.

\subsection{CKKS}

HEAAN (Homomorphic Encryption for Arithmetic of Approximate Numbers) defines a homomorphic encryption (HE) library proposed by Cheon, Kim, Kim and Song (CKKS). The CKKS method uses approximate arithmetics over complex numbers \cite{cheon2017homomorphic}. Overall, it is a levelled approach that involves the evaluation of arbitrary circuits of bounded (pre-determined) depth. These circuits can include ADD (X-OR) and Multiply (AND).

HEAAN uses a rescaling procedure to measure the size of the plaintext. It then produces an approximate rounding due to the truncation of the ciphertext into a smaller modulus. The method is especially useful in that it can be applied to carry out encryption computations in parallel. Unfortunately, the ciphertext modulus can become too small, and where it is not possible to carry out any more operations.

The HEAAN (CKKS) method uses approximate arithmetic over complex numbers ($\mathbb{C}$) and is based on Ring Learning With Errors (RLWE). It focuses on defining an encryption error within the computational error that will happen within approximate computations. We initially take a message (\textit{M}) and convert it to a cipher message (\textit{ct}) using a secret key \textit{sk}. To decrypt ([⟨\textit{ct},\textit{sk}⟩]\textit{q}), we produce an approximate value along with a small error (\textit{e}).

Craig Gentry \cite{gentry2009fully} has outlined three important application areas within privacy-preserving genome association, neural networks, and private information retrieval. Along with this, he proposed that the research community should investigate new methods which did not involve the usage of lattices.

\subsubsection{Chebyshev approximation}
With approximation theory, it is possible to determine an approximate polynomial $p(x)$ that is an approximation to a function $f(x)$. A polynomial takes the form of $p(x)=a_{n}.x^{n}+ a_{n-1}.x^{n-1}+ a_1.x + a_0$, and where $a_0 ... a_n$ are the coefficients of the powers, and $n$ is the maximum power of the polynomial. In this case, we will evaluate arbitrary smooth functions for CKKS and use Chebyshev approximation. These were initially created by Pafnuty Lvovich Chebyshev. This method involves the approximation of a smooth function using polynomials.

Overall, with polynomials, we convert our binary values into a polynomial, such as 101101 is:

\begin{equation}
x^5+x^3+x^2+1
\end{equation}

Our plaintext and ciphertext are then represented as polynomial values.

\subsubsection{Approximation theory}

With approximation theory, we aim to determine an approximate method for a function f(x). It was Pafnuty Lvovich Chebyshev who defined a method of finding a polynomial p(x) that is approximate for f(x). Overall, a polynomial takes the form  of:

\begin{equation}
p(x)=a_n.x^n+a_{n-1}.x^{n-1}+a_1.x+a_0
\end{equation}

and where $a_0 ... a_n$ are the coefficients of the powers, and $n$ is the maximum power of the polynomial. Chebyshev published his work in 1853 as "Theorie des mecanismes, connus sous le nom de parall´elogrammes". His problem statement was “to determine the deviations which one has to add to get an approximated value for a function $f$, given by its expansion in powers of $x-a$, if one wants to minimise the maximum of these errors between $x = a - h$ and $x = a + h$, $h$ being an arbitrarily small quantity".

\subsection{Polynomial evaluations}
A polynomial takes the form form of \(p(x)=a_{n}.x^{n}+ a_{n-1}.x^{n-1}+ a_1.x + a_0\), and where $a_0 ... a_n$ are the coefficients of the powers, and \(n\) is the maximum power of the polynomial. With CKKS in OpenFHE, we can evaluate the result of a polynomial for a given range of $x$ values. For example, if we have $p(x)=5.x^2+3.x+7$ will give a result of $p(2)= 33$.

\section{Related work}

Homomorphic encryption supports the usage of machine learning methods, and some core features include a dot product operation with an encrypted vector and logistic functions. With this, OpenFHE supports a range of relevant methods and even has a demonstrator for a machine learning method.

\subsection{State-of-the-art}
Iezzi et al. \cite{ezzi2020practical} define two methods of training with homomorphic encryption:

\begin{itemize}
    \item Private Prediction as a Service (PPaaS).  This is where the prediction is outsourced to a service provider who has a pre-trained model and where encrypted data is sent to the service provider. In this case, the data owner does not learn the model used.
    \item Private Training as a Service (PTaaS).  This is where the data owner provides data to a service provider and who will train the model. The service provider can then provide a prediction for encrypted data.
\end{itemize}

Wood et al \cite{wood2020homomorphic} adds models of:

\begin{itemize}
    \item Private outsourced computation. This involves moving computation into the cloud.
    \item Private prediction. This involves homomorphic data processed into the cloud, and not having access to the training model. 
    \item Private training. This is where a cloud entity trains a model based on the client's data.
\end{itemize}

\subsection{Basic primitives}
\subsubsection{Logistic Function}
\begin{figure}
\begin{center}
  \includegraphics[width=1.0\linewidth]{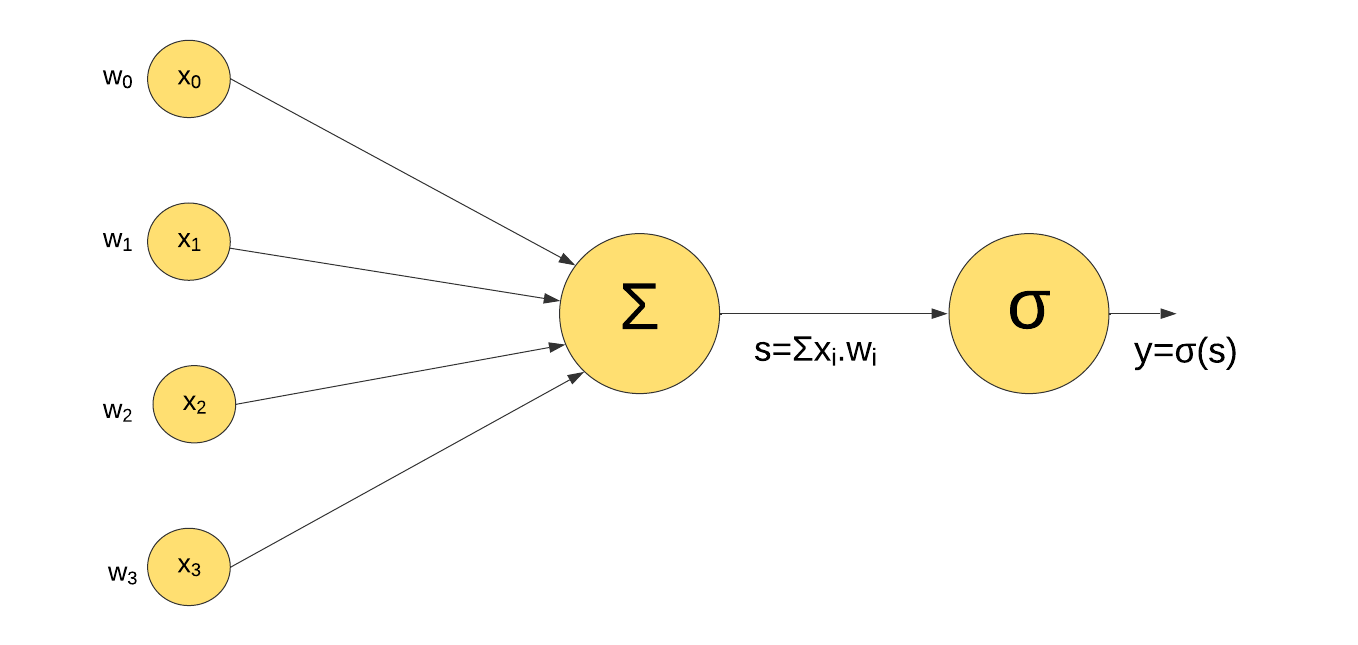}
  \caption{Sigmoid function}
  \label{fig:sigmoid}
\end{center}
\end{figure}

With homomorphic encryption, we can represent a mathematical operation in the form of a homomorphic equation. One of the most widely used methods is to use Chebyshev polynomials, and which allows the mapping of the function to a Chebyshev approximation. A core application of the logistic function - also known as the sigmoid function - is within machine learning. With this, an artificial neural network is created with weighted summation and a sigmoid function (Figure \ref{fig:sigmoid}). Mathematically, this is defined as:

\begin{equation}
f(x)=\frac{1}{1+e^{-x}}
\end{equation}

This is supported in OpenFHE, and which implements Chebyshev approximation. For this, we can use the function of \cite{asecuritysite_42988} and which evaluates 1/(1 + exp(-x)) for f(x), and where $x$ is a range of coefficients with ciphertext. The value of $a$ is the lower bound of the coefficients, and $b$ is the upper bound. The degree value is the desired degree of approximation.

Logistic regression is often used to predict binary outcomes of whether patients need treatment in medical applications, such as with diabetic patients \cite{bender1997ordinal}.

\subsubsection{Inner product}
The inner product of two vectors of $a$ and $b$ is represented by $⟨a,b⟩$. It is the dot product of two vectors and represented as $⟨a,b⟩=|a|.|b|.cos(\theta)$, where $\theta$ is the angle between the two vectors. This operation is supported in OpenFHE \cite{asecuritysite_28805}.

\subsubsection{Matrix operations}
We can perform matrix operations with an encrypted input vector from OpenFHE \cite{asecuritysite_15692}. For this, if we have a vector of the form:

\begin{equation}
v_1 = \begin{bmatrix} x_1 & x_2 & x_3 \end{bmatrix}
\end{equation}

and a matrix of:
\begin{equation}
 m_1 = \begin{bmatrix} w_{11} & w_{21} & w_{31} \\ w_{12} & w_{22} & w_{32} \\ w_{13} & w_{23} & w_{33}  \end{bmatrix}
\end{equation}

We now get:
\begin{equation}
v_1.m_1 = \begin{bmatrix} x_1 & x_2 & x_3 \end{bmatrix} \begin{bmatrix}  w_{11} & w_{21} & w_{31} \\ w_{12} & w_{22} & w_{32} \\ w_{13} & w_{23} & w_{33}  \end{bmatrix}
\end{equation}

and:

\begin{equation}
v_1.m_1 = \begin{bmatrix} x_1 .w_{11} + x_2 .w_{21} + x_3 .w_{31}  & x_1 .w_{12} + x_2 .w_{22} + x_3 .w_{32} & x_1 .w_{13} + x_2 .w_{23} + x_3 .w_{33} \end{bmatrix}
\end{equation}

Thus we get:
\begin{align}
y_1= x_1 .w_{11} + x_2 .w_{21} + x_3 .w_{31} 
y_2= x_1 .w_{12} + x_2 .w_{22} + x_3 .w_{32}
y_3= x_1 .w_{13} + x_2 .w_{23} + x_3 .w_{33} 
 \end{align}
 
\begin{figure}
\begin{center}

  \includegraphics[width=0.8\linewidth]{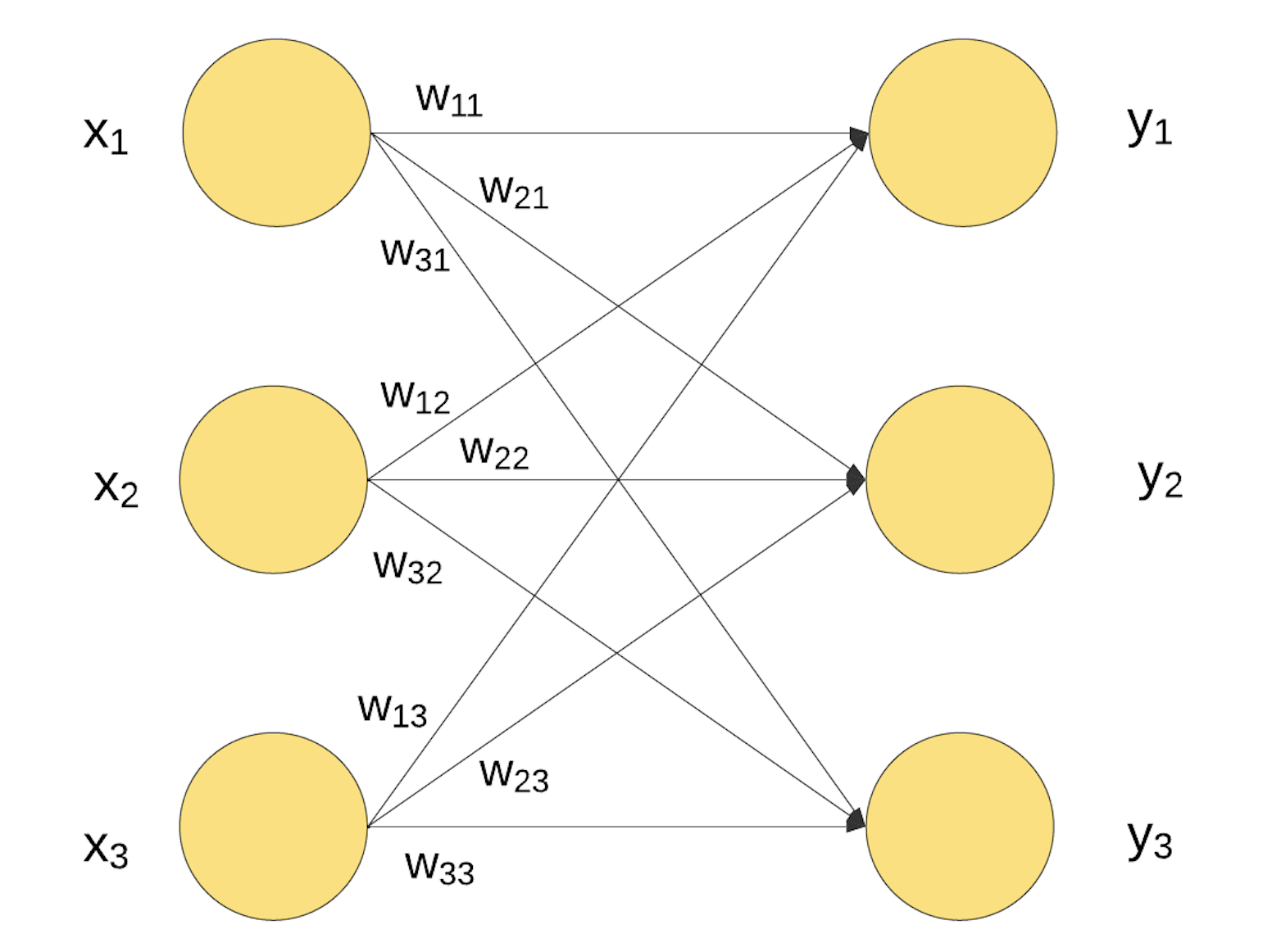}
 \caption{Neural network}
  \label{fig:nn01}
\end{center}
\end{figure}

Figure \ref{fig:nn01} shows this setup.

\subsection{GWAS}
Blatt et al. \cite{blatt2020secure} implemented the Genome-wide association study (GWAS) and which is a secure large-scale genome-wide association study using homomorphic encryption.

\subsubsection{Chi-Square GWAS}
The Chi-squared GWAS test has been implemented in OpenFHE \href{https://github.com/openfheorg/openfhe-genomic-examples/blob/main/demo-chi2.cpp}{Here}. With this, each of the participants in the student group is given a public key from a  GWAS (Genome-wide association studies) coordinator, who then encrypts the data with CKKS and sends it back for processing. The computation includes association statistics using full logistic regression on each variant with sex, age, and age squared as covariates. Pearson’s chi-square test uses categories to determine if there is a significant difference between sets of data \footnote{It implements as RunChi2 from https://github.com/openfheorg/openfhe-genomic-examples/blob/main/demo-chi2.cpp}.

\begin{equation}
\tilde{\chi}^2=\frac{1}{d}\sum_{k=1}^{n} \frac{(O_k - E_k)^2}{E_k}
\end{equation}

and where:

\begin{itemize}
    \item $\tilde{\chi}^2$ is the chi-square test statistic.
    \item $O$ is the observed frequency.
    \item $E$ is the expected frequency.
\end{itemize}

Overall, the implementation involved a dataset of 25,000 individuals, and it was shown that 100,000 individuals and 500,000 single-nucleotide polymorphisms (SNPs) could be evaluated in 5.6~hours on a single server \cite{blatt2020secure}. 

\subsubsection{Linear Regression}
The GWAS method is also implemented with linear regression for homomorphic encryption (See RunLogReg in \href{https://github.com/openfheorg/openfhe-genomic-examples/blob/main/demo-logistic.cpp}{Here}). The results show that the accuracy of both the Chi-squared and linear regression tests was good. The run time varied linearly with the number of participants in the test.

\subsection{Support Vector Machines (SVM)}
With the SVM (Support Vector Machine) model, we have a supervised learning technique. Overall, it is used to create two categories (binary) or more (multi) and will try to allocate each of the training values into one or more categories. Basically, we have points in a multidimensional space and try to create a clear gap between the categories. New values are then placed within one of the two categories.

\begin{figure}
\begin{center}
  \includegraphics[width=1.0\linewidth]{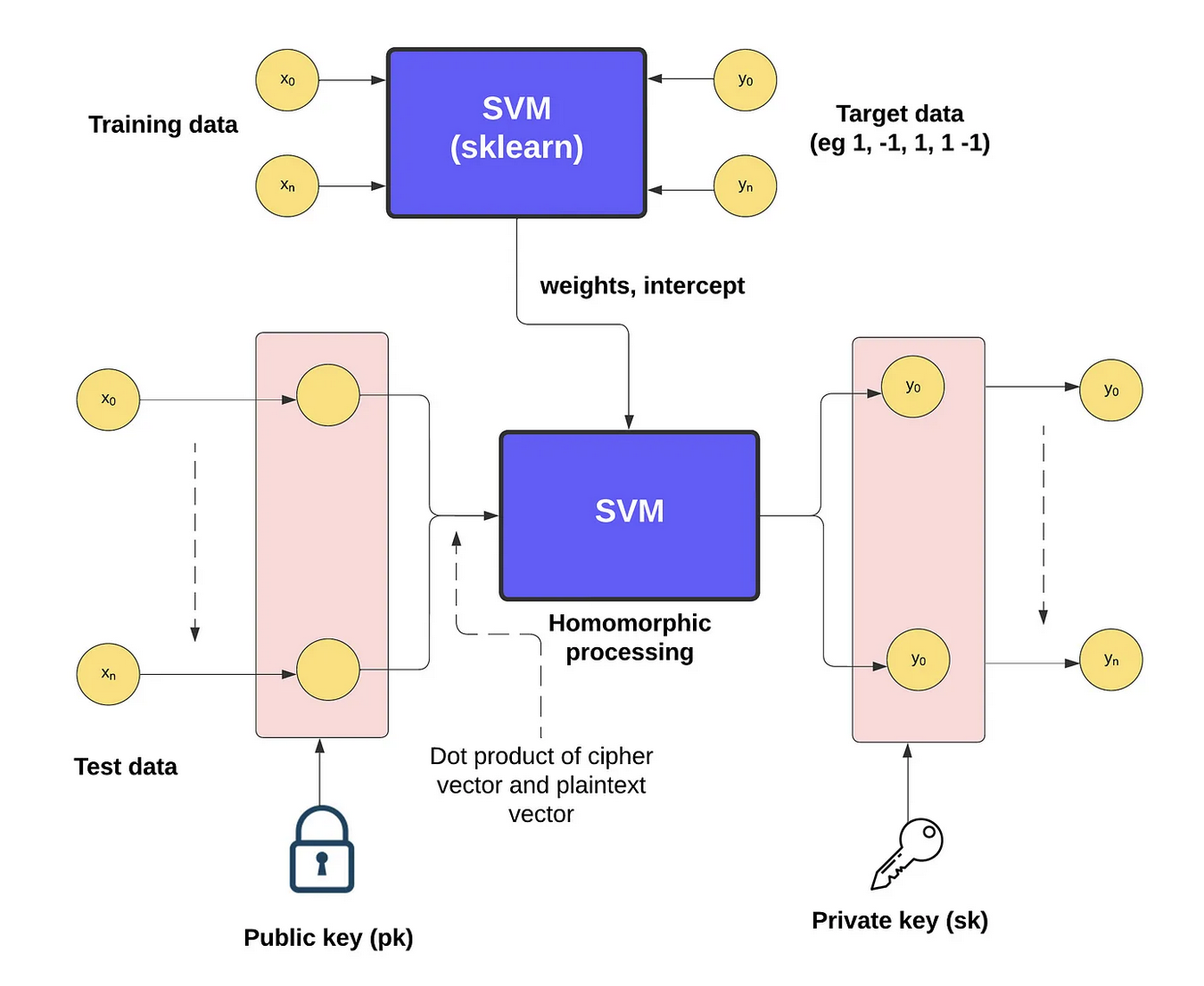}
  \caption{SVM}
 \label{fig:svm}
 \end{center}
\end{figure}

Overall, we split out the input data into training and test data and then train with a sklearn model with unencrypted values from the training data. The output from the model is the weights and intercepts. Next, we can encrypt the test data with the homomorphic public key and then feed this into the SVM model. The output values can then be decrypted by the associated private key, as illustrated in Figure \ref{fig:svm}.

\subsubsection{CKKS and SVM}

The CKKS scheme is a homomorphic encryption method designed for encrypted arithmetic operations. For a given plaintext feature vector of:

\begin{align}
\mathbf{x} = (x_1, x_2, \dots, x_n)
\end{align}

and a public key of $\mathsf{pk}$, the encryption function is:

\begin{equation}
\mathsf{Enc}(\mathbf{x}, \mathsf{pk}) = c_x
    \label{eq:ckks_encryption}
\end{equation}

and where $c_x$ is the encrypted representation of $\mathbf{x}$ \cite{cheon2017homomorphic}. For Support Vector Machine (SVM) classification with Linear SVM, we use a \textbf{linear decision function} of:

\begin{equation}
    f_{\text{lin}}(\mathbf{x}) = \mathbf{w}^T \mathbf{x} + b
    \label{eq:svm_linear_decision}
\end{equation}

and where $\mathbf{x}$ is the feature vector, $\mathbf{w}$ is the weight vector, and $b$ is the bias term.

For classification:

\begin{equation}
    y = \text{sign}(f_{\text{lin}}(\mathbf{x}))
    \label{eq:svm_classification}
\end{equation}

Using FHE, the computation is performed on encrypted values \cite{cheon2017homomorphic}:

\begin{equation}
    \mathsf{Enc}(f_{\text{lin}}(\mathbf{x})) = \mathsf{Enc}(\mathbf{w}^T \mathbf{x} + b)
    \label{eq:homomorphic_linear_svm}
\end{equation}

A Polynomial Kernel SVM extends the decision function to:

\begin{equation}
    f_{\text{poly}}(\mathbf{x}) = (\mathbf{w}^T \mathbf{x} + b)^d
    \label{eq:svm_polynomial_decision}
\end{equation}

where $d$ is the polynomial degree, and where the rest of the parameters are the same as the Linear SVM. For classification:

\begin{equation}
    y = \text{sign}(f_{\text{poly}}(\mathbf{x}))
    \label{eq:svm_polynomial_classification}
\end{equation}

With homomorphic encryption, we  compute this function without decrypting:

\begin{equation}
    \mathsf{Enc}(f_{\text{poly}}(\mathbf{x})) = \mathsf{Enc}((\mathbf{w}^T \mathbf{x} + b)^d)
    \label{eq:homomorphic_polynomial_svm}
\end{equation}

and which follows prior encryption-based SVM work \cite{gentry2009fully,cheon2017homomorphic}. After computation, the result is decrypted using the $\textbf{private key}$ ($\mathsf{sk}$:

\begin{equation}
    \mathsf{Dec}(c_f, \mathsf{sk}) = f(\mathbf{x})
    \label{eq:ckks_decryption}
\end{equation}

The final classification is:

\begin{equation}
    y = \text{sign}(\mathsf{Dec}(c_f, \mathsf{sk}))
    \label{eq:final_classification}
\end{equation}

\section{Methodology}

This paper explores the integration of Fully Homomorphic Encryption (FHE) with Support Vector Machines (SVM) for privacy-preserving machine learning. The proposed framework employs the CKKS encryption scheme, implemented via the OpenFHE library, to enable encrypted inference while maintaining classification accuracy. Homomorphic encryption allows computations to be performed directly on encrypted data without decryption, ensuring data privacy throughout the machine-learning pipeline \cite{cheon2017homomorphic}. OpenFHE is an open-source library that provides implementations of lattice-based encryption schemes, including CKKS, which supports approximate arithmetic operations on encrypted data \cite{openfhe2023}.  

The dataset used for this experiment is the Iris dataset, which contains measurements of iris flowers, including sepal length, sepal width, petal length, and petal width \cite{nguyen2023advancing}. This dataset, originally introduced by Fisher \cite{fisher1936use}, is widely used in machine learning research due to its simplicity and well-separated class distributions. The objective of using this dataset is to evaluate the effectiveness of the encrypted SVM framework in classifying iris species while ensuring data privacy. The dataset is publicly available from the UCI ML Repository \cite{uci2023}.  

The methodology consists of environment setup and data preprocessing. The SVM model, originally introduced by Cortes and Vapnik \cite{cortes1995support}, is trained on plaintext data before being used for encrypted inference. To analyse the trade-offs between encryption security, computational efficiency, and model accuracy, the implementation is conducted using \textit{OpenFHE} within a Python-based machine-learning pipeline, leveraging libraries such as Scikit-Learn and NumPy \cite{geron2019hands}.

\subsection{Environment Setup}  

The encryption parameters in the Fully Homomorphic Encryption (FHE) framework are essential for balancing security, computational efficiency, and model accuracy. The following section outlines the installation process, provides a detailed explanation of each parameter, and presents the system specifications, as shown in Table~\ref{tab:experimental-setup}.

\subsubsection{Experimental Setup}  

The implementation of the encrypted SVM framework followed a structured approach to ensure efficiency and reproducibility. The setup commenced with the installation of \textit{openfhe-python}, adhering strictly to the official guidelines \cite{openfhe_github}. This library provided the essential cryptographic primitives required for executing encrypted computations securely.  

Following the installation, the model training phase was conducted using the \textit{model\_training.py} script. This process involved training an SVM classifier and saving the learned parameters, which were subsequently utilised for encrypted inference. The trained model served as the foundation for performing secure classification without exposing sensitive data.  

To facilitate a standardised evaluation, the dataset was organised within the \textit{data/} directory. If necessary, the dataset could be regenerated by executing the \textit{get\_data.py} script, ensuring consistency and reproducibility across experiments.  

For encrypted inference, two dedicated scripts were employed: \textit{encrypted\_svm\_
linear.py} and \textit{encrypted\_svm\_poly.py}. These scripts enabled inference using linear and polynomial kernel SVM models, respectively, allowing for a comprehensive assessment of encrypted classification performance under different kernel settings. Through this structured approach, the framework effectively demonstrated the feasibility of privacy-preserving machine learning using homomorphic encryption.

\subsubsection{Encryption Parameters}

Homomorphic encryption relies on several key parameters that impact computational efficiency, security, and accuracy \cite{cheon2017homomorphic}. The primary encryption parameters used in this study are:

\begin{itemize}
    \item \textbf{Ring Dimension (\(N\))}: Defines the size of the polynomial ring used in encryption. A larger \(N\) increases security but also raises computational cost \cite{brakerski2014leveled}. Typical values include \(N = 2^{10}, 2^{12}, 2^{14}, \dots\).
    
    \item \textbf{Multiplication Depth (\(D\))}: Represents the number of consecutive multiplications a ciphertext can undergo before noise accumulation becomes a limiting factor \cite{gentry2009fully}. Higher \(D\) enables more complex computations, which is essential for polynomial kernel approximation in SVM.

    \item \textbf{Scaling Factor (\(S\))}: Determines the precision of fixed-point arithmetic in CKKS encryption. A higher \(S\) improves numerical accuracy but increases computational complexity \cite{cheon2018numerical}.

    \item \textbf{First Modulus Size (\(M\))}: Defines the initial modulus size, impacting ciphertext precision and computational overhead \cite{albrecht2018homomorphic}.

    \item \textbf{Security Level (\(L\))}: Specifies the cryptographic strength of encryption (e.g., 128-bit, 192-bit, 256-bit security). A higher \(L\) enhances security but introduces additional computational costs \cite{kim2018batch}.

    \item \textbf{Batch Size (\(B\))}: Represents the number of encrypted values processed in parallel. A larger \(B\) improves computational efficiency, particularly for batch inference \cite{cheon2018numerical}.
\end{itemize}

These parameters significantly impact the feasibility of encrypted machine learning. In our experiments, we analyse their influence on classification accuracy, encryption overhead, and inference efficiency.

\subsubsection{System Specifications}

The system was deployed on an AWS EC2 t3.medium instance, equipped with two virtual CPUs (Intel Xeon 3.1~GHz) and 4~GB of RAM, providing a balanced environment for machine learning and encrypted computations. For software, Python was used as the primary programming language, enabling seamless integration between machine learning and encryption frameworks. scikit-learn then supports the training and evaluation of traditional SVM models, ensuring a robust baseline for comparison. Meanwhile, OpenFHE is used to perform encryption, ciphertext operations, and homomorphic inference, thus enabling secure computation without compromising model performance.

\begin{table}
\centering
\caption{Experimental Setup}
\label{tab:experimental-setup}
\renewcommand{\arraystretch}{1.2}
\begin{tabular}{l l}
\hline
\textbf{Component} & \textbf{Description} \\
\hline
\textbf{Compute Environment} & AWS EC2 \texttt{t3.medium} (Two vCPUs, Intel Xeon 3.1~GHz, \\ & 4~GB RAM) \\
\textbf{Operating System} & Ubuntu 20.04 \\
\textbf{Programming Language} & Python 3.x \\
\textbf{ML Library} & \texttt{scikit-learn} (for SVM training and evaluation) \cite{pedregosa2011scikit} \\
\textbf{HE Library} & OpenFHE (CKKS scheme for encrypted inference) \cite{openfhe2023} \\
\textbf{Dataset} & Iris Dataset (150 samples, four features) \cite{fisher1936use} \\
\textbf{Preprocessing} & Standardisation (zero mean, unit variance), Train-Test\\ & Split (80\%-20\%) \\
\textbf{Encryption Parameters} & \( N, D, S, M, L, B \) (Ring Dim, Mult Depth, Scaling Factor,\\ & Modulus Size, Sec Level, Batch Size) \\
\textbf{SVM Models} & Linear SVM, Polynomial SVM (homomorphic kernel \\ & approximation) \cite{cortes1995support} \\
\textbf{Performance Metrics} & Classification Accuracy, Encryption Overhead, Inference \\ & Time, Memory Usage, Scalability \\
\hline
\end{tabular}
\end{table}

\subsection{Data Preprocessing}

Data preprocessing is a crucial step to ensure reliable and efficient machine learning, particularly when incorporating Homomorphic Encryption \cite{pinto2018iris}. In this study, we use the Iris dataset (150 samples, four features) and apply standardisation to achieve zero mean and unit variance, improving model stability. The data is then split into 80\% training and 20\% testing to enable fair evaluation. Given the constraints of encrypted computation, categorical features are appropriately encoded. These steps help maintain accuracy while minimising computational overhead in secure inference.  

\subsubsection{Dataset Overview}

The Iris dataset is a widely recognised benchmark in machine learning, frequently employed for evaluating classification algorithms \cite{pedregosa2011scikit}. It provides a structured framework for distinguishing between different iris flower species based on their physical attributes. The dataset consists of 150 samples, each representing an individual iris flower, and is characterised by four key features: sepal length, sepal width, petal length, and petal width, all measured in centimetres. These features enable effective classification by capturing the morphological differences among species.  

The dataset comprises three distinct classes, each containing 50 samples, corresponding to three species: \textit{Iris setosa} (label `0`), \textit{Iris versicolor} (label `1`), and \textit{Iris virginica} (label `2`). It is well-structured, balanced, and contains no missing values, making it particularly suitable for both educational purposes and experimental evaluations in machine learning research.  

Due to its simplicity and interpretability, the Iris dataset is commonly used for demonstrating data preprocessing techniques, exploratory data analysis, and classification models, including Support Vector Machines (SVM) and decision trees \cite{cortes1995support}. Furthermore, its features can be visualised through pair plots, enabling an intuitive understanding of feature relationships and class separability.  

In this study, the Iris dataset serves as a controlled environment for analysing the effects of homomorphic encryption on SVM classification. By leveraging its structured nature, we facilitate a reliable comparison between traditional and encrypted inference methods, allowing for a comprehensive assessment of computational performance and classification accuracy.

The Iris dataset is a classic benchmark for machine learning algorithms, favored for its simplicity and accessibility. It is widely used in educational settings and can be easily accessed through libraries like \texttt{scikit-learn} in Python.

\subsubsection{Data Preprocessing and Feature Encoding}

To ensure robust and efficient encrypted classification, the data set was subjected to a systematic preprocessing pipeline implemented by  \texttt{get\_data.py}. This process involved feature selection, transformation, and structuring to optimise the data for FHE-based machine learning.  

The dataset contains 150 rows (samples) and four columns (four predictive features): sepal length, sepal width, petal length, petal width. There is one target column of species classification. Standardisation was applied to normalise values, ensuring a mean ($\mu$) of approximately zero, and a standard deviation ($\sigma$) of approximately unity. This improves model performance by placing features on a similar scale. The data were split into 120 training samples and 30 test samples, ensuring a well-balanced split for model evaluation. Furthermore, categorical labels were encoded into numerical representations to facilitate seamless integration into the machine learning framework.  

This preprocessing stage establishes a structured and standardised foundation for encrypted SVM training. Harmonising feature distributions, optimising data representation, and preparing the dataset for secure computation enhance both the accuracy and efficiency of privacy-preserving machine learning.

\section{Implementation}

This section presents the approach used to implement privacy-preserving classification using FHE. The Support Vector Machine (SVM) model is adapted to operate on encrypted data using the CKKS encryption scheme \cite{cheon2017homomorphic}.

\begin{figure*}
    \centering
    \includegraphics[width=1.1\textwidth]{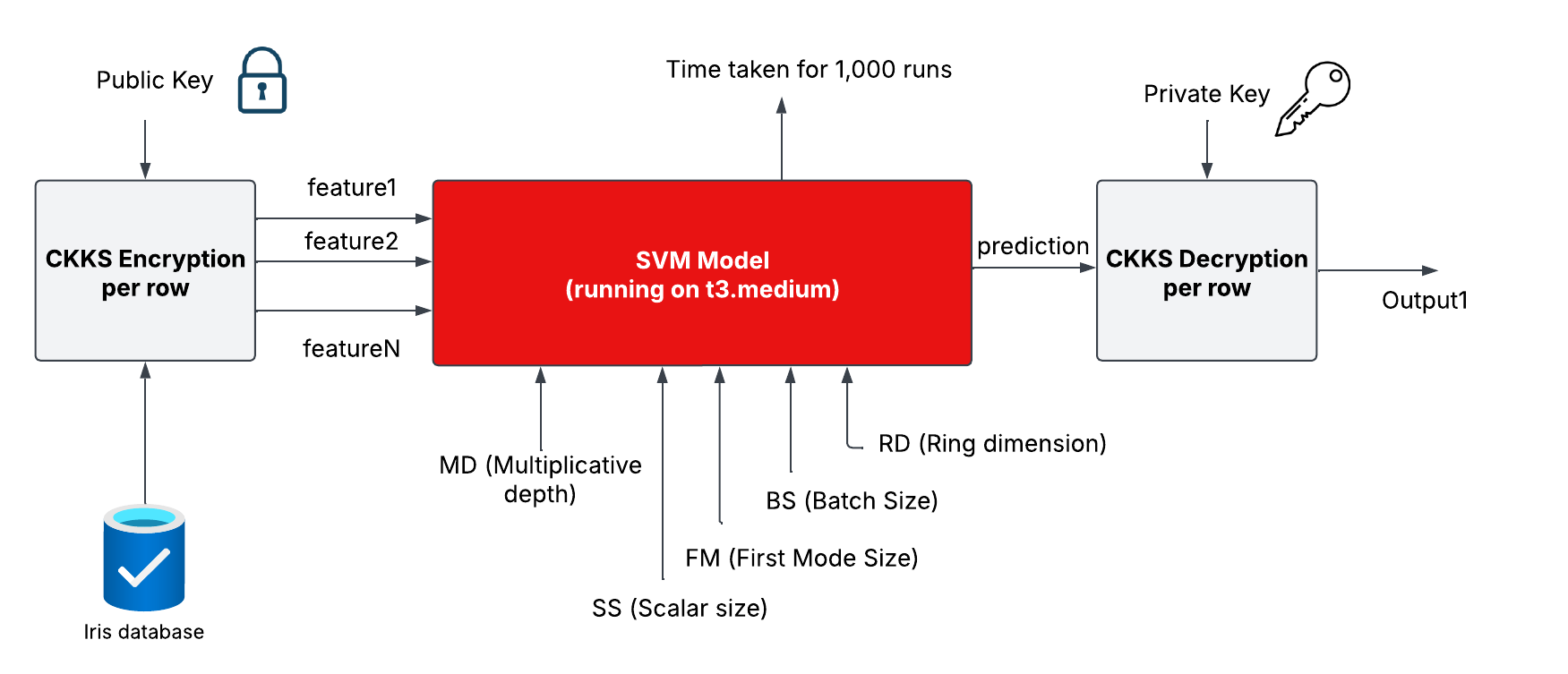}
    \caption{Experimental setup for encrypted classification. The pipeline includes data encryption, encrypted inference, and decryption of results.}
    \label{fig:exp_setup}
\end{figure*}

The implementation consists of dataset preprocessing, encryption of feature vectors, SVM training, and encrypted classification. The process follows established principles from privacy-preserving machine learning \cite{bourse2018fast}. The overall workflow is visualised in Figure \ref{fig:exp_setup}.

The evaluation of the impact of homomorphic encryption on machine learning performance involves a number of experiments measuring key performance metrics. Classification accuracy was assessed by comparing encrypted and non-encrypted inference, with the SVM model achieving high accuracy. Computation time was also analysed, including encryption, inference, and decryption durations. Additionally, the scale-up runtime was examined by calculating the ratio of non-encrypted to encrypted execution times.  

Memory overhead was evaluated to determine the effect of homomorphic encryption on resource consumption, particularly memory usage. Finally, scalability was assessed by analysing performance variations as the ring dimension size and multiplication depth increased.

\subsection {Encryption Using CKKS}

The CKKS encryption scheme was employed to encrypt feature vectors, allowing privacy-preserving computations on floating-point values \cite{cheon2017homomorphic}. CKKS supports approximate arithmetic operations, making it well-suited for machine learning applications. The encryption parameters used in our experiments were selected based on a balance between computational efficiency and security. The multiplicative depth (\(D\)) ranged from 1 to 7, scaling factor (\(S\)) values varied between 10 and 50, and the first modulus size (\(M\)) was tested at 20, 30, 40, 50, and 60. The security level (\(L\)) was evaluated at 128-bit, 192-bit and 256-bit configurations. Batch sizes (\(B\)) included 128, 256, 512, 1024, 2048, and 4096. The ring dimension (\(N\)) was tested at \(2^{14}\) (16,384), \(2^{15}\) (32,768), \(2^{16}\) (65,536), and \(2^{17}\) (131,072), providing insights into the scalability of homomorphic encryption in machine learning.

\subsection{Encrypted Classification Algorithm}
In this work, we propose an FHE-based approach for SVM classification that supports both linear and polynomial kernels. The classification process involves encrypting the feature vector and model parameters, performing homomorphic computations to evaluate the decision function, and decrypting the result to obtain the classification outcome. The detailed steps are outlined in Algorithm~\ref{alg:homomorphic_svm}, which describes the encrypted inference procedure for both linear and polynomial SVM models.

\begin{algorithm}
\caption{\textbf{Homomorphic SVM Classification} \cite{gentry2009fully,cheon2017homomorphic}}
\label{alg:homomorphic_svm}
\begin{algorithmic}[1]
\Require Feature vector \( \mathbf{x} \), public key \( \mathsf{pk} \), private key \( \mathsf{sk} \), degree \( d \) (for polynomial SVM)
\Ensure Classification result \( y \)

\State \textbf{Encrypt Features:} \( c_x \gets \mathsf{Enc}(\mathbf{x}, \mathsf{pk}) \) (Equation~\ref{eq:ckks_encryption})
\State \textbf{Encrypt Model Parameters:} 
\State \quad \( c_w \gets \mathsf{Enc}(\mathbf{w}, \mathsf{pk}) \)
\State \quad \( c_b \gets \mathsf{Enc}(b, \mathsf{pk}) \)
\State \textbf{Compute Encrypted Decision Function:}
\If {Linear SVM}
    \State \( c_f \gets \mathsf{Enc}(\mathbf{w}^T \mathbf{x} + b) \) (Equation~\ref{eq:homomorphic_linear_svm})
\Else {Polynomial SVM}
    \State \( c_f \gets \mathsf{Enc}((\mathbf{w}^T \mathbf{x} + b)^d) \) (Equation~\ref{eq:homomorphic_polynomial_svm})
\EndIf
\State \textbf{Decrypt the Result:} \( f(\mathbf{x}) \gets \mathsf{Dec}(c_f, \mathsf{sk}) \) (Equation~\ref{eq:ckks_decryption})
\State \textbf{Classify Output:} 
\[
y \gets \begin{cases} 
1, & f(\mathbf{x}) \geq 0 \\
-1, & f(\mathbf{x}) < 0
\end{cases} \quad \text{(Equation~\ref{eq:final_classification})}
\]

\State \Return \( y \)

\end{algorithmic}
\end{algorithm}

\subsection{Model training}
The model training example is shown in Appendix B.

\label{sec:evaluation}

\section{Results}  

The experimental results provide a comprehensive evaluation of the impact of homomorphic encryption on SVM inference. A key observation is the trade-off between encryption depth and computational efficiency, where higher security parameters lead to increased execution time and memory consumption. This behaviour is consistent with the theoretical complexity of homomorphic encryption, which introduces overhead due to polynomial arithmetic and ciphertext expansion.  

Beyond computational cost, the study examines the extent to which encrypted inference preserves classification accuracy. By systematically tuning encryption parameters, the analysis explores the balance between security and performance, offering insights into optimising privacy-preserving machine learning. The following sections present a detailed discussion of these findings, grounded in both empirical observations and theoretical considerations.  

Tables \ref{table:results_svm_linear} and \ref{table:results_svm_Poly} data gathered. MD is multiplicative depth, SS is scalar size, FM is first mod size, BS is batch size and RD is the ring dimension. AEA is Average Encryption Accuracy, NEA is Non-Encrypted Accuracy, AET is Average Encryption Time, and ANT is Average Non-Encryption Time.

\begin{table}
\centering
\setlength{\tabcolsep}{5pt} 
\begin{tabular}{c c c c c c | c c c c c}
\hline
MD & SS	& FM	& SL	& BS	& RD	& AEA & NEA & AET & ANT & Scale up\\
\hline
1	& 30	& 60	& 128& 	1,024& 	16,384& 	0.967 & 	0.967 & 0.643458 & 	0.000623 & 1,032.838\\
2 & 30 & 60 & 128 & 1,024 & 16,384 & 0.967 & 0.967 & 0.782 & 0.000067 & 1,172.735\\
3 & 30 & 60 & 128 & 1,024 & 16,384 & 0.967 & 0.967 & 0.924 & 0.000161 & 1,397.215\\
4 & 30 & 60 & 128 & 1,024 & 16,384 & 0.967 & 0.967 & 1.101 & 0.000613 & 1,794.548\\
5 & 30 & 60 & 128 & 1,024 & 16,384 & 0.967 & 0.967 & 1.283 & 0.000624 & 2,056.393\\
6 & 30 & 60 & 128 & 1,024 & 16,384 & 0.967 & 0.967 & 1.391 & 0.000627 & 2,097.537\\
7 & 30 & 60 & 128 & 1,024 & 16,384 & 0.967 & 0.967 & 1.530 & 0.000658 & 2,324.503\\
\hline
1 & 10 & 60 & 128 & 1,024 & 16,384 & 0.817 & 0.967 & 0.649 & 0.000067 & 9,697.24\\
1 & 20 & 60 & 128 & 1,024 & 16,384 & 0.967 & 0.967 & 0.672 & 0.000065 & 10,332.49\\
1 & 30 & 60 & 128 & 1,024 & 16,384 & 0.967 & 0.967 & 0.651 & 0.000073 & 8,919.69\\
1 & 40 & 60 & 128 & 1,024 & 16,384 & 0.967 & 0.967 & 0.650 & 0.000029 & 22,431.52\\
1 & 50 & 60 & 128 & 1,024 & 16,384 & 0.967 & 0.967 & 0.650 & 0.000027 & 24,084.56\\
\hline
1 & 30 & 20 & 128 & 1,024 & 16,384 & 0.967 & 0.967 & 0.627 & 0.000076 & 8,251.12\\
1 & 30 & 30 & 128 & 1,024 & 16,384 & 0.967 & 0.967 & 0.632 & 0.000067 & 9,434.24\\
1 & 30 & 40 & 128 & 1,024 & 16,384 & 0.967 & 0.967 & 0.635 & 0.000074 & 8,573.4\\
1 & 30 & 50 & 128 & 1,024 & 16,384 & 0.967 & 0.967 & 0.643 & 0.000065 & 9,905.05\\
1 & 30 & 60 & 128 & 1,024 & 16,384 & 0.967 & 0.967 & 0.641 & 0.000067 & 9,574.09\\
\hline
1 & 30 & 60 & 192 & 1,024 & 16,384 & 0.967 & 0.967 & 0.204 & 0.000184 & 1,108.59\\
1 & 30 & 60 & 256 & 1,024 & 16,384 & 0.967 & 0.967 & 0.197 & 0.000188 & 1,050.52\\
1 & 30 & 60 & 512 & 1,024 & 16,384 & 0.967 & 0.967 & 0.201 & 0.000191 & 1,050.31\\
1 & 30 & 60 & 1,024 & 1,024 & 16,384 & 0.967 & 0.967 & 0.198 & 0.000193 & 1,023.06\\
1 & 30 & 60 & 2,048 & 1,024 & 16,384 & 0.967 & 0.967 & 0.202 & 0.000193 & 1,048.5\\
1 & 30 & 60 & 4,096 & 1,024 & 16,384 & 0.967 & 0.967 & 0.647 & 0.000711 & 910.78\\
\hline
1 & 30 & 60 & 128 & 128 & 16,384 & 0.967 & 0.967 & 0.198 & 0.000109 & 1,817.4\\
1 & 30 & 60 & 128 & 256 & 16,384 & 0.967 & 0.967 & 0.195 & 0.000107 & 1,822.5\\
1 & 30 & 60 & 128 & 512 & 16,384 & 0.967 & 0.967 & 0.196 & 0.000109 & 1,808.7\\
1 & 30 & 60 & 128 & 1,024 & 16,384 & 0.967 & 0.967 & 0.641 & 0.000067 & 9,574.88\\
1 & 30 & 60 & 128 & 2,048 & 16,384 & 0.967 & 0.967 & 0.206 & 0.000109 & 1,878.6\\
1 & 30 & 60 & 128 & 4,096 & 16,384 & 0.967 & 0.967 & 0.213 & 0.000109 & 1,945.3\\
\hline
1 & 30 & 60 & 128 & 1,024 & 16,384 & 0.967 & 0.967 & 0.638 & 0.000662 & 963.82\\
1 & 30 & 60 & 128 & 1,024 & 32,768 & 0.967 & 0.967 & 1.265 & 0.000624 & 2,026.52\\
1 & 30 & 60 & 128 & 1,024 & 65,536 & 0.967 & 0.967 & 2.583 & 0.00067 & 3,646.68\\
1 & 30 & 60 & 128 & 1,024 & 131,072 & 0.967 & 0.967 & 5.103 & 0.00065 & 8,245.34\\
\hline
\end{tabular}
   \caption{Results for SVM-Linear}
    \label{table:results_svm_Poly}
\end{table}

\begin{table}
\centering
\setlength{\tabcolsep}{5pt} 
\begin{tabular}{c c c c c c | c c c c c}
\hline
MD & SS	& FM	& SL	& BS	& RD	& AEA & NEA & AET & ANT & Scale up\\
\hline

1 & 30 & 60 & 128 & 1,024 & 16,384 & 0.967 & 0.967 & 0.648 & 0.000708 & 915.2\\
2 & 30 & 60 & 128 & 1,024 & 16,384 & 0.967 & 0.967 & 0.783 & 0.000730 & 1,093.3\\
3 & 30 & 60 & 128 & 1,024 & 16,384 & 0.967 & 0.967 & 0.919 & 0.000763 & 1,405.0\\
4 & 30 & 60 & 128 & 1,024 & 16,384 & 0.967 & 0.967 & 1.097 & 0.000629 & 1,527.7\\
5 & 30 & 60 & 128 & 1,024 & 16,384 & 0.967 & 0.967 & 1.288& 0.000773 & 1,665.7\\
6 & 30 & 60 & 128 & 1,024 & 16,384 & 0.967 & 0.967 & 1.399 & 0.000712 & 1,954.9\\
7 & 30 & 60 & 128 & 1,024 & 16,384 & 0.967 & 0.967 & 1.562 & 0.000715 & 2,248.4\\
\hline
1 & 10 & 60 & 128 & 1,024 & 16,384 & 0.784 & 0.967 & 0.649 & 0.000067 & 973.8\\
1 & 20 & 60 & 128 & 1,024 & 16,384 & 0.967 & 0.967 & 0.671 & 0.000065 & 1,032.4\\
1 & 30 & 60 & 128 & 1,024 & 16,384 & 0.967 & 0.967 & 0.651& 0.000073 & 889.4\\
1 & 40 & 60 & 128 & 1,024 & 16,384 & 0.967 & 0.967 & 0.650 & 0.000029 & 1,033.9\\
1 & 50 & 60 & 128 & 1,024 & 16,384 & 0.967 & 0.967 & 0.650 & 0.000027 & 1,036.4\\
\hline
1 & 30 & 20 & 128 & 1,024 & 16,384 & 0.967 & 0.967 & 0.627 & 0.000076 & 927.7\\
1 & 30 & 30 & 128 & 1,024 & 16,384 & 0.967 & 0.967 & 0.632 & 0.000067 & 940.1\\
1 & 30 & 40 & 128 & 1,024 & 16,384 & 0.967 & 0.967 & 0.634 & 0.000074 & 941.8\\
1 & 30 & 50 & 128 & 1,024 & 16,384 & 0.967 & 0.967 & 0.643& 0.000065 & 998.1\\
1 & 30 & 60 & 128 & 1,024 & 16,384 & 0.967 & 0.967 & 0.641 & 0.000067 & 961.3\\
\hline
1 & 30 & 60 & 192 & 1,024 & 16,384 & 0.967 & 0.967 & 0.203 & 0.000184 & 1,090.9\\
1 & 30 & 60 & 256 & 1,024 & 16,384 & 0.967 & 0.967 & 0.197 & 0.000188 & 1,049.6\\
1 & 30 & 60 & 512 & 1,024 & 16,384 & 0.967 & 0.967 & 0.201& 0.000191 & 1,052.2\\
1 & 30 & 60 & 1,024 & 1,024 & 16,384 & 0.967 & 0.967 & 0.197 & 0.000193 & 1,023.7\\
1 & 30 & 60 & 2,048 & 1,024 & 16,384 & 0.967 & 0.967 & 0.202 & 0.000193 & 1,048.9\\
1 & 30 & 60 & 4,096 & 1,024 & 16,384 & 0.967 & 0.967 & 0.646 & 0.000711 & 908.7\\
\hline
1 & 30 & 60 & 128 & 128   & 16,384 & 0.967 & 0.967 & 0.641 & 0.000067 & 961.3\\
1 & 30 & 60 & 128 & 256   & 16,384 & 0.967 & 0.967 & 0.204 & 0.000184 & 1,090.9\\
1 & 30 & 60 & 128 & 512   & 16,384 & 0.967 & 0.967 & 0.197& 0.000188 & 1,049.6\\
1 & 30 & 60 & 128 & 1,024  & 16,384 & 0.967 & 0.967 & 0.201 & 0.000191 & 1,052.2\\
1 & 30 & 60 & 128 & 2,048  & 16,384 & 0.967 & 0.967 & 0.198 & 0.000193 & 1,023.7\\
1 & 30 & 60 & 128 & 4,096  & 16,384 & 0.967 & 0.967 & 0.202 & 0.000193 & 1,048.9\\
\hline
1 & 30 & 60 & 128 & 1,024 & 16,384 & 0.967 & 0.967 & 0.680 & 0.000747 & 910.6\\
1 & 30 & 60 & 128 & 1,024 & 32,768 & 0.967 & 0.967 & 1.269 & 0.000668 & 1,951.5\\
1 & 30 & 60 & 128 & 1,024 & 65,536 & 0.967 & 0.967 & 2.549 & 0.000604 & 3,935.1\\
1 & 30 & 60 & 128 & 1,024 & 131,072 & 0.967 & 0.967 & 5.245 & 0.000687 & 7,716.4\\
\hline
\end{tabular}
\caption{Results for SVM-poly}
\label{table:results_svm_linear}
\end{table}

\subsubsection{Classification Accuracy}
Table~\ref{tab:accuracy_comparison} presents the classification accuracy of plaintext and encrypted SVM models. The results show that, in this experiment, homomorphic encryption has no significant impact on model accuracy, as both versions achieve similar performance. This confirms the effectiveness of the CKKS encryption scheme in preserving the integrity of machine learning inference.

\begin{table}
\centering
\caption{Classification Accuracy Comparison}
\label{tab:accuracy_comparison}
\renewcommand{\arraystretch}{1.2}
\begin{tabular}{l l}
\hline
\textbf{Model} & \textbf{Accuracy (\%)} \\
\hline
\textbf{SVM (Plaintext)} & 96.7 \\
\textbf{SVM (Encrypted)} & 96.7 \\
\hline
\end{tabular}
\end{table}

\subsubsection{Computational Overhead}
Homomorphic encryption introduces additional computational costs due to encryption, encrypted inference, and decryption steps. Table~\ref{tab:runtime_analysis} compares execution times for plaintext and encrypted models.

\begin{table}
\centering
\caption{Runtime Analysis (sec)}
\label{tab:runtime_analysis}
\renewcommand{\arraystretch}{1.2}
\begin{tabular}{l c c}
\hline
\textbf{Operation} & \textbf{Plaintext} & \textbf{Encrypted} \\
\hline
Feature Encryption & - & 0.2029 \\
Inference & 0.0002 & 0.2029 \\
Decryption & - & 0.0001 \\
\hline
\end{tabular}
\end{table}

The encrypted inference process is around 1,000 times slower than plaintext execution, primarily due to polynomial evaluations performed under encryption.

\subsubsection{Scalability and Resource Utilisation}

Scalability ensures stable performance as data grows, while resource utilisation optimises computational efficiency. Balancing encryption parameters helps maintain security, accuracy, and performance.

\subsubsection{Impact of Ring Dimension Size}
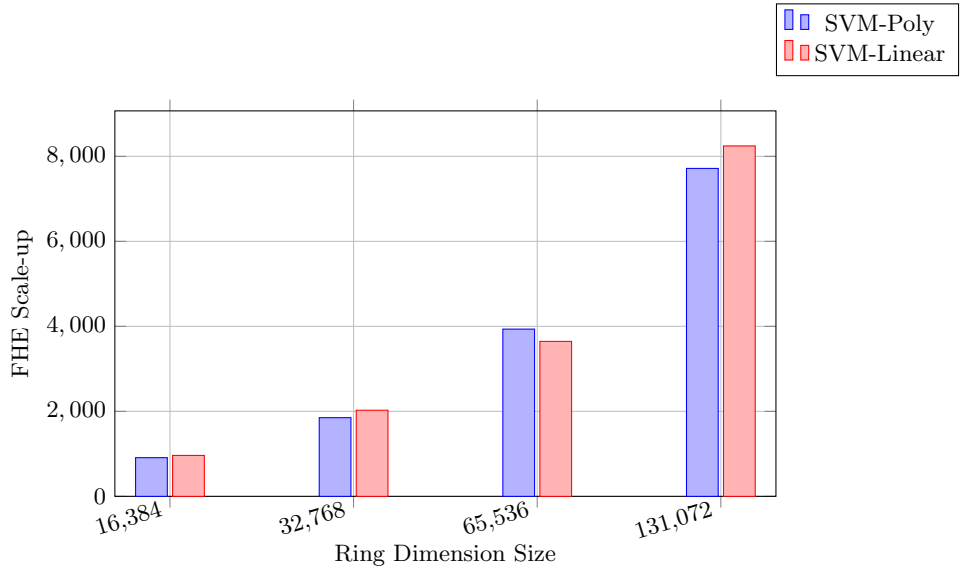
\begin{figure}[]
    \centering
    \begin{tikzpicture}
        \begin{axis}[
            ybar,
            ylabel={FHE Scale-up},
            xlabel={Ring Dimension Size},
            symbolic x coords={16384, 32768, 65536, 131072},
            xtick=data,
            ymin=0,
            legend style={at={(1,1.09)}, anchor=south west},
            width=0.85\textwidth,
            height=0.55\textwidth,
            grid=major,
            bar width=12pt,
            xticklabel style={rotate=17, anchor=east},
            xticklabels={{16,384}, {32,768}, {65,536}, {131,072}},
            yticklabel style={/pgf/number format/.cd, 1000 sep={,}} 
        ]
        \addplot coordinates {(16384,910) (32768,1851) (65536,3935) (131072,7716)};
        \addplot coordinates {(16384,963) (32768,2026) (65536,3646) (131072,8245)};
        \legend{SVM-Poly, SVM-Linear}
        \end{axis}
    \end{tikzpicture}
    \caption{Homomorphic Scale-up vs SVM-Linear and SVM-Poly Ring Dimension}
    \label{fig:bar_chart}
\end{figure}

The effect of increasing the ring dimension on encrypted inference time is shown in Table~\ref{tab:ring_dimension_results} and Figure~\ref{fig:bar_chart}. Larger ring dimensions increase computation time due to expanded ciphertext size.

\begin{table}
\centering
\small 
\renewcommand{\arraystretch}{1} 
\setlength{\tabcolsep}{2.3pt} 
\caption{Homomorphic Scale-up with Varying Ring Dimensions}
\label{tab:ring_dimension_results}
\begin{tabular}{c c c c c c | c c}
\hline
\textbf{MD} & \textbf{SS} & \textbf{FM} & \textbf{SL} & \textbf{BS} & \textbf{RD} & \textbf{SVM-Linear} & \textbf{SVM-Poly} \\
\hline
1 & 30 & 60 & 128 & 1024 & 16K  & 963.8  & 910.6  \\
1 & 30 & 60 & 128 & 1024 & 32K  & 2,026.5 & 1,851.7 \\
1 & 30 & 60 & 128 & 1024 & 64K  & 3,646.7 & 3,935.1 \\
1 & 30 & 60 & 128 & 1024 & 128K & 8,245.3 & 7,716.4 \\
\hline
\end{tabular}
\end{table}

\subsubsection{Impact of Multiplication Depth on FHE Scale-up}
\begin{figure}
    \centering
    \begin{tikzpicture}
        \begin{axis}[
            ybar,
            ylabel={FHE Scale-up},
            xlabel={MultDepth Range},
            xtick={1,2,3,4,5,6,7},
            xticklabels={1,2,3,4,5,6,7},
            ymin=0,
            legend style={at={(1,1.09)}, anchor=south west},
            width=0.85\textwidth,
            height=0.55\textwidth,
            grid=major,
            bar width=12pt,
            symbolic x coords={1,2,3,4,5,6,7},
            xticklabel style={rotate=0, anchor=east}
        ]
        \addplot coordinates {(1,915.2) (2,1093.3) (3,1460.5) (4,1527.7) (5,1665.7) (6,1954.9) (7,2248.4)};
        \addplot coordinates {(1,1032.8) (2,1172.7) (3,1397.2) (4,1794.5) (5,2056.4) (6,2097.5) (7,2324.5)};
        \legend{SVM-Poly, SVM-Linear}
        \end{axis}
    \end{tikzpicture}
    \caption{Homomorphic Scale-up vs SVM-Linear and SVM-Poly MultDepth}
    \label{fig:bar_chart1}
\end{figure}
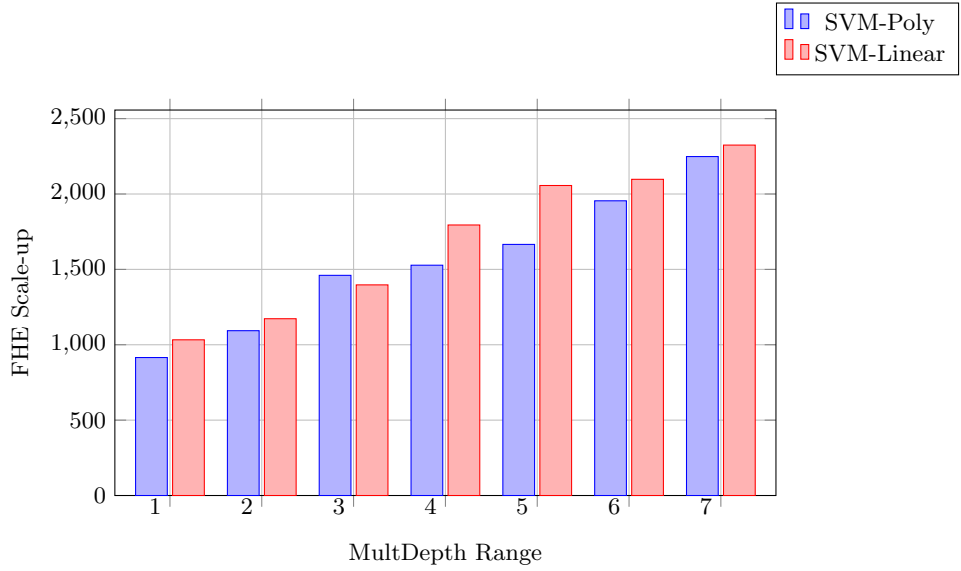

Table~\ref{tab:multipath_runtime2} presents the effect of increasing multiplication depth \(D\) on encrypted inference speed, as visualised in Figure~\ref{fig:bar_chart1}.

\begin{table}
\centering

\setlength{\tabcolsep}{5pt} 
\renewcommand{\arraystretch}{1.2} 
\caption{Homomorphic Scale-up for SVM-Linear and SVM-Poly with Varying Multiplication Depth}
\label{tab:multipath_runtime2}

\begin{tabular}{c c c c c c | c c}
\hline
\textbf{MD} & \textbf{SS} & \textbf{FM} & \textbf{SL} & \textbf{BS} & \textbf{RD} & \textbf{SVM-Linear} & \textbf{SVM-Poly} \\
\hline
1 & 30 & 60 & 128 & 1024 & 16,384  & 1,032.8 &  915.2  \\
2 & 30 & 60 & 128 & 1024 & 16,384  & 1,172.7 & 1,093.3  \\
3 & 30 & 60 & 128 & 1024 & 16,384  & 1,397.2 & 1,460.5  \\
4 & 30 & 60 & 128 & 1024 & 16,384  & 1,794.5 & 1,527.7  \\
5 & 30 & 60 & 128 & 1024 & 16,384  & 2,056.4 & 1,665.7  \\
6 & 30 & 60 & 128 & 1024 & 16,384  & 2,097.5 & 1,954.9  \\
7 & 30 & 60 & 128 & 1024 & 16,384  & 2,324.5 & 2,248.4  \\
\hline
\end{tabular}
\end{table}

The experimental results highlight key trade-offs in homomorphic encryption for SVM inference. Table~\ref{tab:accuracy_comparison} confirms that the encrypted SVM model maintains 96.7\% accuracy, similar to the plaintext model, demonstrating that CKKS encryption does not affect classification performance. Table~\ref{tab:runtime_analysis} shows that encrypted inference is approximately 1,000 times slower than plaintext inference, primarily due to polynomial evaluations under encryption. This slowdown arises from the computational complexity of homomorphic operations, particularly ciphertext multiplication and relinearisation~\cite{cheon2017homomorphic}. Unlike plaintext arithmetic, where multiplication is a constant-time operation, homomorphic multiplication involves modular reductions, rescaling, and key-switching, leading to significant overhead \cite{halevi2014algorithms}.

\subsubsection{Impact of Multiplication Depth} 

Table~\ref{tab:multipath_runtime2} and Figure~\ref{fig:bar_chart1} reveal that increasing multiplication depth significantly raises execution time, with SVM-Poly scale-up increasing from 915.2s at depth 1 to 2248.4s at depth 7. The reason is that multiplication depth determines the number of sequential homomorphic multiplications that can be performed before bootstrapping is required~\cite{gentry2009fully}. As the depth increases: 

\begin{itemize}
    \item Noise Growth. Each multiplication amplifies noise, requiring frequent relinearisation and rescaling, which are computationally expensive~\cite{cheon2017homomorphic}.
    \item 
    \item Exponentially Larger Ciphertexts. Higher-depth computations require larger ciphertext modulus values to maintain correctness, increasing memory and computation costs~\cite{sealmanual}.
    \item 
    \item Bootstrapping Overhead. If the noise exceeds the threshold, a bootstrapping step is needed, which further increases execution time~\cite{halevi2014algorithms,buchanan2025partial}.
\end{itemize}

These findings emphasise the need for optimisation strategies, including ciphertext packing, bootstrapping, and hardware acceleration, to improve the feasibility of encrypted machine learning in real-world applications.  

\subsubsection{Limitations and Future Work}  

While the proposed approach demonstrates promising results, certain limitations must be addressed to enhance its practical applicability. The most significant challenge lies in the high computational cost of homomorphic encryption, which leads to substantial execution time overhead. This limitation poses a significant barrier to real-time applications, particularly in scenarios where rapid inference is required. Furthermore, the large memory footprint associated with ciphertext storage presents scalability concerns, especially for deployment on resource-constrained devices.  

The increased computational burden can be attributed to the underlying complexity of homomorphic encryption operations, which involve polynomial arithmetic over large integer rings \cite{cheon2017homomorphic}. The reliance on number-theoretic transforms (NTTs) for polynomial multiplication introduces an inherent O(n.log n) computational cost, while the quadratic complexity of matrix-vector operations within SVM classification further compounds execution time \cite{gentry2009fully}. Additionally, the trade-off between multiplication depth and accuracy, dictated by the hardness of the Ring Learning With Errors (RLWE) problem, influences both performance and security \cite{brakerski2014leveled}.  

To mitigate these challenges, future research should explore hardware acceleration techniques, such as leveraging GPUs and FPGAs, to enhance computational efficiency \cite{dai2019accelerating}. Additionally, optimising encryption parameters—such as ring dimension size and coefficient modulus selection—can significantly reduce latency and memory consumption \cite{cheon2017homomorphic}. Exploring alternative cryptographic schemes, such as hybrid encryption approaches (Typically, it merges a fast symmetric encryption scheme with a secure asymmetric encryption scheme to balance efficiency and security), may further improve the feasibility of encrypted machine learning in real-world applications \cite{mo2022efficient}. 

\section{Conclusion} 
The usage of homomorphic encryption within machine learning provides great hope for privacy-aware learning. Unfortunately, it will come with an overhead of processing. This paper shows that using an extracted SVM model provides an excellent method of creating a model which can then be used to process data. In order to understand the key parameters which affect performance, the paper evaluates multiplication depth, scale size, first modulus size, security level, batch size, and ring dimension, along with two different SVM models, SVM-Poly and SVM-Linear. Overall, the results show that the two main parameters which affect performance are the ring dimension and the modulus size, and that SVM-Poly and SVM-Linear show similar performance levels.




\section{Appendix A}
In applying our SVM implementation, we can use sklearn to train the model \href{https://github.com/openfheorg/python-svm-examples/blob/master/model_training.py}{Here}:

\begin{python}
import pandas as pd
import numpy as np
from sklearn.svm import SVC

# Load the data
X_train = pd.read_csv('data/credit_approval_train.csv')
X_test = pd.read_csv('data/credit_approval_test.csv')
y_train = pd.read_csv('data/credit_approval_target_train.csv')
y_test = pd.read_csv('data/credit_approval_target_test.csv')

# Model Training
print("---- Starting Models Training ----")

print("Starting SVM Linear")
svc_linear = SVC(kernel='linear')
svc_linear.fit(X_train, y_train.values.ravel())
print("SVM Linear Completed")

svc_poly = SVC(kernel='poly',degree=3,gamma=2)
svc_poly.fit(X_train, y_train.values.ravel())
print("SVM Poly Completed")

print("---- Model Training Completed! ----")

decision_function = svc_linear.decision_function(X_test)
ytestscore = decision_function[0]

decision_function_poly = svc_poly.decision_function(X_test)
ytestscore_poly = decision_function_poly[0]

# Saving Results
np.savetxt("models/weights.txt", svc_linear.coef_)
np.savetxt("models/intercept.txt", svc_linear.intercept_)
np.savetxt("data/ytestscore.txt", [ytestscore])
np.savetxt("models/dual_coef.txt", svc_poly.dual_coef_)
np.savetxt("models/support_vectors.txt", svc_poly.support_vectors_)
np.savetxt("models/intercept_poly.txt", svc_poly.intercept_)
np.savetxt("data/ytestscore_poly.txt", [ytestscore_poly])
\end{python}

This splits the input data into training and test data. The training data is then used to train the model with a linear and a polynomial SVM training model. It then outputs the model with a number of weights and intercept values. Next, we can run our homomorphic encryption method and take the training data (x), the weights, and the bias for processing [here]:

\begin{python}
pt_x = cc.MakeCKKSPackedPlaintext(x)
pt_weights = cc.MakeCKKSPackedPlaintext(weights.tolist())
pt_bias = cc.MakeCKKSPackedPlaintext([intercept])
\end{python}

These values remain as plaintext values. We can then encrypt the training data with the public key [here]:

\begin{python}
ct_x = cc.Encrypt(keys.publicKey, pt_x)
\end{python}

We then create an inner product with the cipher training data and the weights [here]:

\begin{python}
ct_res = cc.EvalInnerProduct(ct_x, pt_weights,n)
\end{python}

An example of implementing a dot product is here. The output is then the multiplication (inner product) of the cipher values of the training data and the weights. Next, we can mask out the first value with:

\begin{python}
mask = [0] * n
mask[0] = 1
pt_mask = cc.MakeCKKSPackedPlaintext(mask)
ct_res = cc.EvalMult(ct_res, pt_mask)
\end{python}

Then we add the bias:
\begin{python}
ct_res = cc.EvalAdd(ct_res, pt_bias)
\end{python}

Finally, we can decrypt the resultant value with the private key:
\begin{python}
result = cc.Decrypt(ct_res, keys.secretKey)
\end{python}

\section{Appendix B}
The model training process was executed using the following command to train and save the model weights. The encrypted model files were subsequently used for inference:

\begin{verbatim}
python model_training.py
\end{verbatim}

Upon execution, the following output was produced:

\begin{verbatim}
---- Starting Models Training ----
Starting SVM Linear
SVM Linear Completed
Starting SVM Poly
SVM Poly Completed
---- Model Training Completed! ----
All results saved successfully!
\end{verbatim}

The dataset required for training is located in the data/ directory. However, to regenerate the dataset, the following command was executed:

\begin{verbatim}
python get_data.py
\end{verbatim}

The script selected the following features:
\small
\begin{verbatim}
Total number of features in dataset: 4
Selected: ['sepal length (cm)', 'sepal width (cm)', 'petal length (cm)', 'petal width (cm)']\end{verbatim}
\normalsize

The original dataset contained 150 samples with 4 features:

\small
\begin{verbatim}
Original data shape: (150, 4)
Features: ['sepal length (cm)', 'sepal width (cm)', 'petal length (cm)', 'petal width (cm)']\end{verbatim}
\normalsize

Standardisation was applied, producing the following results:

\begin{verbatim}
Standardisation results:
sepal length (cm): mean=-0.000, std=1.003
sepal width (cm): mean=-0.000, std=1.003
petal length (cm): mean=-0.000, std=1.003
petal width (cm): mean=-0.000, std=1.003
\end{verbatim}

The dataset was then split into training and testing sets:

\begin{verbatim}
Data processing completed successfully!

Saved files:
Training samples: 120
Testing samples: 30
Number of selected features: 4
\end{verbatim}

These experiments provide insights into the trade-offs between security and computational efficiency in privacy-preserving machine learning. The execution of encrypted inference using 
\begin{verbatim}
encrypted_svm_linear.py 
\end{verbatim}
produced the following output:

\begin{verbatim}
---- Testing OpenFHE Encryption ----
Original data: [1.5, 2.0, 3.5]
Decrypted values (real part): [1.4999997346263885, 1.9999997120806627, 3.499999428471009]
---- Running Encrypted Linear SVM ----
Avg Encrypted SVM Accuracy: 0.9667
Avg Non-Encrypted SVM Accuracy: 0.9667
Avg Encrypted Time: 0.2029 sec
Avg Non-Encrypted Time: 0.0002 sec
\end{verbatim}

\bibliographystyle{IEEEtran}
\bibliography{main}

\begin{thebibliography}{10}
\providecommand{\url}[1]{#1}
\csname url@samestyle\endcsname
\providecommand{\newblock}{\relax}
\providecommand{\bibinfo}[2]{#2}
\providecommand{\BIBentrySTDinterwordspacing}{\spaceskip=0pt\relax}
\providecommand{\BIBentryALTinterwordstretchfactor}{4}
\providecommand{\BIBentryALTinterwordspacing}{\spaceskip=\fontdimen2\font plus
\BIBentryALTinterwordstretchfactor\fontdimen3\font minus \fontdimen4\font\relax}
\providecommand{\BIBforeignlanguage}[2]{{%
\expandafter\ifx\csname l@#1\endcsname\relax
\typeout{** WARNING: IEEEtran.bst: No hyphenation pattern has been}%
\typeout{** loaded for the language `#1'. Using the pattern for}%
\typeout{** the default language instead.}%
\else
\language=\csname l@#1\endcsname
\fi
#2}}
\providecommand{\BIBdecl}{\relax}
\BIBdecl

\bibitem{openfhe_github}
O.~D. Team, ``Openfhe: Open-source fully homomorphic encryption library,'' GitHub Repository, 2023, \url{https://github.com/openfheorg/openfhe-development}.

\bibitem{rivest1978data}
R.~L. Rivest, L.~Adleman, M.~L. Dertouzos \emph{et~al.}, ``On data banks and privacy homomorphisms,'' \emph{Foundations of secure computation}, vol.~4, no.~11, pp. 169--180, 1978.

\bibitem{asecuritysite_17070}
\BIBentryALTinterwordspacing
W.~J. Buchanan, ``Openfhe,'' \url{https://github.com/openfheorg/openfhe-development}, OpenFHE, 2024, accessed: Feb 20, 2025. [Online]. Available: \url{https://github.com/openfheorg/openfhe-development}
\BIBentrySTDinterwordspacing

\bibitem{homenc}
C.~Gentry, ``A fully homomorphic encryption scheme,'' 2009, \url{crypto.stanford.edu/craig}.

\bibitem{van2010fully}
M.~Van~Dijk, C.~Gentry, S.~Halevi, and V.~Vaikuntanathan, ``Fully homomorphic encryption over the integers,'' in \emph{Advances in Cryptology--EUROCRYPT 2010: 29th Annual International Conference on the Theory and Applications of Cryptographic Techniques, French Riviera, May 30--June 3, 2010. Proceedings 29}.\hskip 1em plus 0.5em minus 0.4em\relax Springer, 2010, pp. 24--43.

\bibitem{brakerski2014efficient}
Z.~Brakerski and V.~Vaikuntanathan, ``Efficient fully homomorphic encryption from (standard) lwe,'' \emph{SIAM Journal on computing}, vol.~43, no.~2, pp. 831--871, 2014.

\bibitem{cheon2017homomorphic}
J.~H. Cheon, A.~Kim, M.~Kim, and Y.~Song, ``Homomorphic encryption for arithmetic of approximate numbers,'' in \emph{Advances in Cryptology--ASIACRYPT 2017: 23rd International Conference on the Theory and Applications of Cryptology and Information Security, Hong Kong, China, December 3-7, 2017, Proceedings, Part I 23}.\hskip 1em plus 0.5em minus 0.4em\relax Springer, 2017, pp. 409--437.

\bibitem{asecuritysite_85691}
\BIBentryALTinterwordspacing
W.~J. Buchanan, ``Homomorphic encryption (seal),'' \url{https://asecuritysite.com/seal}, Asecuritysite.com, 2024, accessed: September 04, 2024. [Online]. Available: \url{https://asecuritysite.com/seal}
\BIBentrySTDinterwordspacing

\bibitem{asecuritysite_40933}
\BIBentryALTinterwordspacing
------, ``Homomorphic encryption with bfv using node.js,'' \url{https://asecuritysite.com/seal/js_homomorphic}, Asecuritysite.com, 2025, accessed: February 28, 2025. [Online]. Available: \url{https://asecuritysite.com/seal/js_homomorphic}
\BIBentrySTDinterwordspacing

\bibitem{wood2020homomorphic}
A.~Wood, K.~Najarian, and D.~Kahrobaei, ``Homomorphic encryption for machine learning in medicine and bioinformatics,'' \emph{ACM Computing Surveys (CSUR)}, vol.~53, no.~4, pp. 1--35, 2020.

\bibitem{ducas2015fhew}
L.~Ducas and D.~Micciancio, ``Fhew: bootstrapping homomorphic encryption in less than a second,'' in \emph{Annual international conference on the theory and applications of cryptographic techniques}.\hskip 1em plus 0.5em minus 0.4em\relax Springer, 2015, pp. 617--640.

\bibitem{al2023demystifying}
A.~Al~Badawi and Y.~Polyakov, ``Demystifying bootstrapping in fully homomorphic encryption,'' \emph{Cryptology ePrint Archive}, 2023.

\bibitem{asecuritysite_65179}
\BIBentryALTinterwordspacing
W.~J. Buchanan, ``Chebyshev approximations using openfhe and c++ (logarithm methods),'' \url{https://asecuritysite.com/openfhe/openfhe_18cpp}, Asecuritysite.com, 2024, accessed: September 04, 2024. [Online]. Available: \url{https://asecuritysite.com/openfhe/openfhe_18cpp}
\BIBentrySTDinterwordspacing

\bibitem{gentry2009fully}
C.~Gentry, ``Fully homomorphic encryption using ideal lattices,'' in \emph{Proceedings of the forty-first annual ACM symposium on Theory of computing}, 2009, pp. 169--178.

\bibitem{ezzi2020practical}
M.~Iezzi, ``Practical privacy-preserving data science with homomorphic encryption: an overview,'' in \emph{2020 IEEE International Conference on Big Data (Big Data)}.\hskip 1em plus 0.5em minus 0.4em\relax IEEE, 2020, pp. 3979--3988.

\bibitem{asecuritysite_42988}
\BIBentryALTinterwordspacing
W.~J. Buchanan, ``Logistic (sigmoid) function evaluation using openfhe and c++,'' \url{https://asecuritysite.com/openfhe/openfhe_20cpp}, Asecuritysite.com, 2024, accessed: September 06, 2024. [Online]. Available: \url{https://asecuritysite.com/openfhe/openfhe_20cpp}
\BIBentrySTDinterwordspacing

\bibitem{bender1997ordinal}
R.~Bender and U.~Grouven, ``Ordinal logistic regression in medical research,'' \emph{Journal of the Royal College of physicians of London}, vol.~31, no.~5, p. 546, 1997.

\bibitem{asecuritysite_28805}
\BIBentryALTinterwordspacing
W.~J. Buchanan, ``Ckks inner product using openfhe and c++,'' \url{https://asecuritysite.com/openfhe/openfhe_13cpp}, Asecuritysite.com, 2024, accessed: September 05, 2024. [Online]. Available: \url{https://asecuritysite.com/openfhe/openfhe_13cpp}
\BIBentrySTDinterwordspacing

\bibitem{asecuritysite_15692}
\BIBentryALTinterwordspacing
------, ``Matrix multiplication with homomomorphic encryption for bfv using openfhe and c++,'' \url{https://asecuritysite.com/openfhe/openfhe_26cpp}, Asecuritysite.com, 2024, accessed: September 06, 2024. [Online]. Available: \url{https://asecuritysite.com/openfhe/openfhe_26cpp}
\BIBentrySTDinterwordspacing

\bibitem{blatt2020secure}
M.~Blatt, A.~Gusev, Y.~Polyakov, and S.~Goldwasser, ``Secure large-scale genome-wide association studies using homomorphic encryption,'' \emph{Proceedings of the National Academy of Sciences}, vol. 117, no.~21, pp. 11\,608--11\,613, 2020.

\bibitem{openfhe2023}
O.~Contributors, ``Fully homomorphic encryption library,'' 2023, available at \url{https://github.com/openfheorg/openfhe-development}.

\bibitem{nguyen2023advancing}
T.~Nguyen, ``Advancing privacy and accuracy with federated learning and homomorphic encryption,'' \emph{Authorea Preprints}, 2023.

\bibitem{fisher1936use}
R.~A. Fisher, ``The use of multiple measurements in taxonomic problems,'' \emph{Annals of Eugenics}, vol.~7, no.~2, pp. 179--188, 1936.

\bibitem{uci2023}
U.~M.~L. Repository, ``Iris dataset,'' 2023, available at \url{https://archive.ics.uci.edu/ml/datasets/iris}.

\bibitem{cortes1995support}
C.~Cortes and V.~Vapnik, ``Support-vector networks,'' \emph{Machine Learning}, vol.~20, no.~3, pp. 273--297, 1995.

\bibitem{geron2019hands}
A.~Géron, \emph{Hands-On Machine Learning with Scikit-Learn, Keras, and TensorFlow}, 2nd~ed.\hskip 1em plus 0.5em minus 0.4em\relax O'Reilly Media, 2019.

\bibitem{brakerski2014leveled}
Z.~Brakerski, ``Efficient fully homomorphic encryption from (standard) lwe,'' \emph{SIAM Journal on Computing}, vol.~43, no.~2, pp. 831--871, 2014.

\bibitem{cheon2018numerical}
J.~H. Cheon, A.~Kim, M.~Kim, and Y.~Song, ``Numerical methods for homomorphic encryption,'' \emph{Cryptology ePrint Archive}, 2018.

\bibitem{albrecht2018homomorphic}
M.~Albrecht, R.~Player, and S.~Scott, ``On the concrete hardness of learning with errors,'' \emph{Journal of Mathematical Cryptology}, vol.~12, no.~3, pp. 169--203, 2018.

\bibitem{kim2018batch}
A.~Kim, Y.~Song, and J.~H. Cheon, ``Batching methods for homomorphic encryption,'' \emph{Cryptology ePrint Archive}, 2018.

\bibitem{pedregosa2011scikit}
F.~Pedregosa, G.~Varoquaux, and A.~e.~a. Gramfort, ``Scikit-learn: Machine learning in python,'' \emph{Journal of Machine Learning Research}, vol.~12, pp. 2825--2830, 2011.

\bibitem{pinto2018iris}
J.~P. Pinto, S.~Kelur, and J.~Shetty, ``Iris flower species identification using machine learning approach,'' in \emph{2018 4th International Conference for Convergence in Technology (I2CT)}.\hskip 1em plus 0.5em minus 0.4em\relax IEEE, 2018, pp. 1--4.

\bibitem{bourse2018fast}
F.~Bourse, M.~Minelli, M.~Minihold, and P.~Paillier, ``Fast homomorphic evaluation of deep discretized neural networks,'' in \emph{Advances in Cryptology – CRYPTO 2018}, ser. Lecture Notes in Computer Science, vol. 10992.\hskip 1em plus 0.5em minus 0.4em\relax Springer, 2018, pp. 483--512.

\bibitem{halevi2014algorithms}
S.~Halevi and V.~Shoup, ``Algorithms in helib,'' \emph{Advances in Cryptology – CRYPTO 2014}, p. 554–571, 2014.

\bibitem{sealmanual}
M.~S. Team, ``Microsoft seal (simple encrypted arithmetic library),'' 2022, available at \url{https://www.microsoft.com/en-us/research/project/microsoft-seal/}.

\bibitem{buchanan2025partial}
W.~J. Buchanan and H.~Ali, ``Partial and fully homomorphic matching of ip addresses against blacklists for threat analysis,'' \emph{arXiv preprint arXiv:2502.16272}, 2025.

\bibitem{dai2019accelerating}
W.~Dai, B.~Deloingce, H.~Chen, and K.~Laine, ``Accelerating fully homomorphic encryption using gpu,'' \emph{Proceedings of the 26th IEEE International Symposium on High-Performance Computer Architecture (HPCA)}, p. 85–98, 2019.

\bibitem{mo2022efficient}
Y.~Mo, J.~Liu, Y.~Zhang, and K.~Ren, ``Efficient hybrid homomorphic encryption for encrypted machine learning,'' \emph{IEEE Transactions on Dependable and Secure Computing}, vol.~19, no.~3, p. 1601–1614, 2022.

\end{thebibliography}

\end{document}